\documentclass[sigconf,authorversion]{acmart}

\usepackage{float}
\usepackage{tikz}
\usepackage{amsmath}
\usepackage{appendix}
\usepackage{chngcntr}
\usepackage{caption}
\usepackage{subcaption}
\usepackage{makecell}
\usepackage{longtable}
\usepackage{multicol}
\usepackage{caption} 

\AtBeginDocument{%
  \providecommand\BibTeX{{%
    \normalfont B\kern-0.5em{\scshape i\kern-0.25em b}\kern-0.8em\TeX}}}

\copyrightyear{2022}
\acmYear{2022}
\setcopyright{rightsretained}

\acmConference[CCS '22]{Proceedings of the 2022 ACM SIGSAC Conference on Computer and Communications Security}{November 7--11, 2022}{Los Angeles, CA, USA}
\acmBooktitle{Proceedings of the 2022 ACM SIGSAC Conference on Computer and Communications Security (CCS '22), November 7--11, 2022, Los Angeles, CA, USA}
\acmDOI{10.1145/3548606.3560626}
\acmISBN{978-1-4503-9450-5/22/11}

\usepackage{etoolbox}
\makeatletter
\patchcmd{\maketitle}{\@copyrightpermission}{
   \begin{minipage}{0.3\columnwidth}
     \href{https://creativecommons.org/licenses/by/4.0/}{\includegraphics[width=0.90\textwidth]{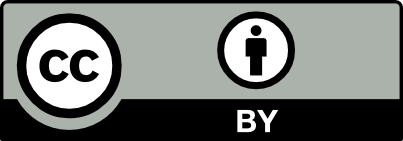}}
   \end{minipage}\hfill
   \begin{minipage}{0.7\columnwidth}
     \href{https://creativecommons.org/licenses/by/4.0/}{This work is licensed under a Creative Commons Attribution International 4.0 License.}
   \end{minipage}

   \vspace{5pt}
}{}{}

\makeatother

\begin{document}

\title{Privacy Limitations of Interest-based Advertising on The Web: A Post-mortem Empirical Analysis of Google's FLoC}

\author{Alex Berke}
\email{aberke@mit.edu}
\affiliation{
    \institution{MIT Media Lab}
    \country{USA}
}

\author{Dan Calacci}
\email{dcalacci@media.mit.edu}
\affiliation{%
    \institution{MIT Media Lab}
    \country{USA}
}

\renewcommand{\shortauthors}{Alex Berke \& Dan Calacci}

\begin{abstract}
In 2020, Google announced it would disable third-party cookies in the Chrome browser to improve user privacy. In order to continue to enable interest-based advertising while mitigating risks of individualized user tracking, Google proposed FLoC. The FLoC algorithm assigns users to "cohorts" that represent groups of users with similar browsing behaviors so that ads can be served to users based on their cohort. In 2022, after testing FLoC in a real world trial, Google canceled the proposal with little explanation in favor of another way to enable interest-based advertising. This work provides a post-mortem analysis of two critical privacy risks for FloC by applying an implementation of FLoC to a real-world browsing history dataset collected from over 90,000 U.S. devices over a one year period.

First, we show how, contrary to its privacy goals, FLoC would have \textit{enabled} individualized cross-site user tracking by providing a unique identifier for users available across sites, similar to the third-party cookies FLoC was meant to be an improvement over. 
We show how FLoC cohort ID sequences observed over time can provide this unique identifier to trackers, even with third-party cookies disabled.
We estimate the number of users in our dataset that could be uniquely identified by FLoC IDs is more than 50\% after 3 weeks and more than 95\% after 4 weeks.
We also show how these risks increase when cohort data are combined with browser fingerprinting, and how our results are conservative underestimates of the risks FLoC would have posed in a real-world deployment. 
Second, we examine the risk of FLoC leaking sensitive demographic information about users. Although we find statistically significant differences in browsing behaviors between demographic groups, we do not find that FLoC significantly risks exposing race or income information about users in our dataset. 
Our contributions provide insights and example analyses for future novel approaches that seek to protect user privacy while monetizing the web.
\end{abstract}



\maketitle

\section{Introduction}

Online advertising is considered a pillar of "free" content on the web~\cite{estrada-jimenezOnlineAdvertisingAnalysis2017}. Advertisers can choose to place specific ads for users based on a variety of information. This information has previously been described under 3 main categories: (1) First-party contextual information about the page a user is viewing, (2) general information about the interests of a user who may be landing on a page where the ad can be served, and (3) specific actions or browsing histories recorded for the user~\cite{FederatedLearningCohorts2021}. In the current advertising ecosystem, (2) and (3) are typically enabled by "third-party" cookies, which can be used to track individual users across the sites they visit~\cite{estrada-jimenezOnlineAdvertisingAnalysis2017, bashirDiffusionUserTracking2018}. This tracking can be used to build profiles for users representing their past internet behaviors, interests, or other information. Privacy advocates have considered this kind of tracking detrimental to user privacy and many web browsers now disable third-party cookies by default. 

Following suit, Google announced it would also remove third-party cookies from the Chrome browser and would provide a new system to enable interest-based advertising while mitigating the risks of individualized user tracking and better protect user privacy. 
In 2020, Google proposed FLoC as such a system \cite{flocwhitepaper2020}. 
FLoC was described as a method to enable ad targeting based on general browsing interests without exposing individual browsing histories~\cite{FederatedLearningCohorts2021}.  In the proposed method, a browser uses the FLoC algorithm to compute a user's "interest cohort" based on their browsing history, where a cohort includes thousands of users with a similar recent browsing history. The cohort ID is then made accessible to advertisers and other third-party trackers or observers.

Following the proposal, Google ran a trial for what it considered a preliminary implementation of FLoC, and millions of Chrome browser users were automatically included in the trial \cite{WhatFLoC}.
At the start of 2022, Google canceled the FLoC proposal in favor of an alternative system, which would again be designed to support interest-based advertising and better preserve user privacy. The change of plans came with little explanation, such as analysis that resulted from the trial. Was FLoC able to provide utility for interest-based advertising? Was FLoC able to sufficiently protect user privacy?
Our analyses address questions about privacy and raise questions about the utility, and potential privacy-utility trade offs, that FLoC would have provided. 

Even before Google completed their FLoC trial, privacy advocates described FLoC as a "misguided" attempt to reinvent behavioral targeting \cite{rescorlaTechnicalCommentsFLoC2021}, and described specific privacy risks enabled by FLoC. For example, a Mozilla report suggested that when considered as a replacement for third-party cookies, FLoC may still enable user tracking \cite{rescorlaTechnicalCommentsFLoC2021}. Others raised the issue that FLoC may reveal sensitive information about individuals based on their cohort, such as their likely demographics \cite{rescorlaTechnicalCommentsFLoC2021, cyphersGoogleFLoCTerrible2021}. However, to our knowledge, no empirical evaluation of these risks has been shown either by Google or external researchers.

\paragraph{Contributions:}
In this paper we provide a post-mortem empirical analysis, evaluating two privacy concerns raised about FLoC: (1) FLoC may enable cross-site user tracking, even with third-party cookies disabled, and (2) FLoC may reveal sensitive demographic information about users.
To evaluate these concerns, we implement the FLoC algorithm used in Google's 2021 trial. We compute cohorts for users in a real-world dataset of browsing history collected over a one-year period from over 90,000 devices owned by 50,000 households distributed across the United States. 

For (1), we show that even with third-party cookies disabled, trackers can use FLoC to identify users across sites.
While FLoC cohorts are $k$-anonymous when considering a given time period in isolation, we show how a user device's sequence of cohort IDs over time can be unique and used to track them across domains. 
More than half of the cohort ID sequences we compute from our dataset are unique after only 3 weeks, and 95\% of them are unique after 4 weeks.
We also show how unicity risks increase when cohort IDs are combined with a weak proxy for browser fingerprinting.

For (2), we do not find that FLoC leaks sensitive demographic information about users in our dataset.
We repurpose a version of \textit{t-closeness} used by Google to identify cohorts that visit sensitive domains at a higher rate than the general population.
Using this metric, we show that the likelihood that FLoC clusters users by race or income is no greater than random chance.
These results are surprising given that we also find browsing behaviors significantly differ by demographic group.

The analysis methods we develop and the significant privacy risks we find can provide useful insights for future proposals that aim to improve user web privacy.

\paragraph{Outline:}
The paper is outlined as follows. The \nameref{section:background} section provides a brief overview of what cookies are, their relationship to online advertising, and the more privacy-preserving alternatives being developed. It also provides a brief technical overview of FLoC in order to give the reader necessary information to understand the privacy risks we address and how our empirical analyses address them. The \nameref{section:background} section also describes these risks. 
The \nameref{section:data and computing cohorts} section then describes our data sources, preprocessing methods, and how cohorts are computed and used throughout the analyses.
Our two main analyses are then presented in separate sections. Section \ref{section:unicity} addresses the risk of individualized user tracking enabled by FLoC.
Section \ref{section:floc sensitivity to race and income} addresses the risk of FLoC leaking sensitive user demographics.
We then provide a discussion of our results and conclusion.

\section{Background}
\label{section:background}

\subsection{Cookies, Interest-based Advertising, And FLoC As A Solution}

Cookies \cite{barthHTTPStateManagement2011} provide a storage mechanism allowing websites to maintain information about users across sessions. They enable many important features of modern web browsing, such as keeping users logged in or maintaining shopping carts. 
When a user visits a website, HTTP requests are made to domains with content embedded on the website, and the cookies are sent in the requests.

There is an important distinction between "first-party" cookies and "third-party" cookies. 

"Third-party" cookies are currently the main mechanism enabling cross-site user tracking. 
They are used by third-party observers or trackers (we will refer to them as "third-parties"), who embed content across many websites in order to track users across the sites they visit.
Consider a third-party with content embedded on many websites, including sites A and B. When a user visits one of these sites (A), the third-party can store a unique identifier for this user in a cookie. When the user visits another site (B), the third-party can access the user identifier set on the first site (A) via a third-party cookie, and see that the user is the same user they tracked on site A.
Tracking users across sites in this way allows ad-tech companies to build profiles for users based on sites they visit, infer their interests, and serve personalized ads. 2018 estimates show that many advertising and analytics companies can observe at least 91\% of an average user’s browsing history, and even with ad blockers companies still observe 40–90\%~\cite{bashirDiffusionUserTracking2018}.

In contrast, "first-party" cookies isolate data by a same-origin policy. This means data stored on one site cannot be accessed on another site. Using first-party cookies alone does not support cross-site user tracking, because there is no mechanism to link data across sites, providing no way to know that a user visiting two sites is the same user.

For this reason, first-party cookies and other storage mechanisms that are limited to a same-origin policy are more privacy-preserving, and are being further developed as ways to enhance utility on the web as third-party cookies are phased out.

Partitioned storage is one such same-origin policy storage mechanism, which is already implemented in the Firefox browser~\cite{StatePartitioningPrivacy} and proposed for Chrome~\cite{privacysandbox}.
When a user visits a particular domain, partitioned storage allows embedded third-parties access to a store unique to that third-party and domain. This "partitions" the storage that third-parties have access to based on the domain a user is visiting.
Like with first-party cookies, this limits a third-party’s ability to track users across sites, as a third-party cannot access the data stored for a user across domains.

However, a third-party may still be able to track a user across sites if they are able to derive the same unique ID for a user across these sites. We describe how storage mechanisms with same-origin policies might be leveraged in combination with FLoC cohort IDs for this purpose in Section~\ref{section:privacy concerns for floc}.

While the ability to track users across sites is a privacy issue, it also plays a large role in the profitability for ad serving ecosystems.
In 2019, the Google Display Ads team ran a randomized controlled experiment to empirically quantify the effect that disabling access to third-party cookies would have on the programmatic ad revenue of web publishers~\cite{ravichandranEffectDisablingThirdparty2019}. They ran the experiment with the ad serving system that is used by Google services to place ads on non-Google sites across the web. They observed that for the top 500 global publishers, average revenue in the treatment group decreased by 52\%, with a median per-publisher decline of 64\%.

\paragraph{FLoC as a solution:}
In order to address privacy concerns without substantially disrupting the advertising industry, Google announced plans to remove third-party cookies from Chrome and proposed FLoC as an alternative means to enable interest-based advertising. At a high level, FLoC aimed to protect privacy by grouping users into k-anonymous \cite{sweeneyKanonymityModelProtecting2002} cohorts. A user's cohort is computed within their browser, based on their browsing history, and sites can then access a user's cohort ID via a browser API. By observing cohort IDs across sites, advertisers can then build profiles about cohorts rather than individuals.\footnote{For a graphical description of how FLoC can be used instead of third-party cookies, see \url{https://web.dev/floc/\#how-does-floc-work.}}

\subsection{How FLoC Works}
FLoC stands for "Federated Learning of Cohorts". However, despite the name, federated learning was not used in the implementation of FLoC. In this section we give a high-level description of the FLoC implementation that was used in Google's real-world trial, which we also implemented for the analysis in this work. More details can be found in Google's documentation~\cite{FLoCOriginTrial}. 

The input to calculate a user's FLoC cohort ID is the set of public domains (eTLD+1's) a user visited in the preceding 7-day period leading up to cohort calculation. The browser uses this set of domain names to produce a hash bitvector using SimHash, which is in a family of locality-sensitive hash (LSH) algorithms \cite{zhaoLocalityPreservingHashing2014}. The featured property of SimHash is that similar vectors are more likely to have the same hash value than dissimilar vectors - meaning similar browsing histories are more likely to result in the same SimHash value.

To enforce k-anonymity, hash values are sorted into groups of size $k$ using Google's PrefixLSH algorithm, described in~\cite{epastoClusteringPrivateInterestbased2021}. While SimHash can be computed locally within the browser, this sorting procedure must have access to all hash values and therefore occur centrally. This sorting is done by what Google refers to as an "anonymity server" which produces a mapping of SimHash value prefixes to cohort IDs.  The mapping is stored within browsers to allow them to locally calculate cohort IDs.

FLoC is designed so that cohort IDs can be recomputed periodically, allowing a user's cohort to change over time as their browsing behavior changes. The frequency with which cohorts are recomputed was not explicitly defined; this work assumes cohorts are recomputed every 7 days.

\subsection{FLoC Trial (Origin Trial)}
FLoC was tested in an origin trial (OT)~\cite{FLoCOriginTrial} from Spring to Fall, 2021 (Chrome 89 to 91) \cite{HowTakePart}. The trial included Chrome users with at least 7 domains in their browsing history. 
The OT FLoC implementation used $k=2000$, resulting in 33,872 cohorts. Approximately 2.3\% of these cohorts were deemed sensitive and dropped, as further described below \cite{FLoCOriginTrial}.
We note that $k$ is a parameter in the FLoC algorithm that may be tuned to a dataset size; both $k$ and dataset size impact the number of cohorts created.

\subsection{Privacy Concerns For FLoC}
\label{section:privacy concerns for floc}

Numerous privacy concerns about FLoC were raised \cite{rescorlaTechnicalCommentsFLoC2021, cyphersGoogleFLoCTerrible2021}. In this section we highlight those addressed in our analysis.

\subsubsection{FLoC May Enable Cross-site User Tracking}

\begin{table*}[ht]
\caption{Toy example illustrating the risk of users being tracked by cohort ID sequences and fingerprinting data. Users' cohorts are recomputed periodically and can change over time. By observing unique cohort ID sequences, third-parties may identify and track users across the web. These privacy risks increase when cohort ID sequences are combined with fingerprinting data.}
\begin{center}
\begin{tabular}{c c c c c c c c}
\hline
& User device 1 & User device 2 & User device 3 & User device 4 & User device 5 & User device 6 \\
\hline
Fingerprinting data & A & A & B & B & C & C \\
Period 1 cohort ID & 1 & 1 & 1 & 2 & 2 & 2 \\
Period 2 cohort ID & 1 & 2 & 2 & 1 & 1 & 2 \\
Period 3 cohort ID & 1 & 2 & 1 & 1 & 2 & 2 \\
\hline 
\end{tabular}
\end{center}

\label{table:cohort_id_sequences}
\end{table*}

Even without third-party cookies, it is still possible for third-parties to track users across sites if they have access to a unique identifier for users that is consistent across the sites they visit.
FLoC could enable cross-site user tracking by providing such an identifier.
This is because the FLoC cohort ID for a user changes over time, and as a result, a user's sequence of cohort IDs may be unique, or shared with only a few other users, breaking the guarantees of $k$-anonymity.

Third-parties who observe users' cohort IDs on the various sites users visit can leverage first-party cookies or partitioned storage to store the cohort IDs for each user on each site. This way they can accumulate a store of the sequence of cohort IDs for each user they observe, and identify unique sequences.
The likelihood of unicity can increase when a user’s cohort ID sequence is combined with fingerprinting data.

Fingerprinting~\cite{eckersleyHowUniqueYour2010} is a separate mechanism from cookies that allows third-party observers to collect information from a user's browser to help identify the user, such as through IP addresses \cite{mishraDonCountMe2020} or device-specific information \cite{gomez-boixHidingCrowdAnalysis2018}. 
While many user devices might share a particular cohort ID, or cohort ID sequence, a much smaller subset of those users might also share a particular browser "fingerprint". 

To better illustrate this privacy risk, we provide a toy example, adapted from a previous description of this risk by Mozilla researchers \cite{rescorlaTechnicalCommentsFLoC2021}. This is shown in Table~\ref{table:cohort_id_sequences}.
This toy example assumes a web browsing universe with only 6 user devices, where each has associated fingerprinting data.
Cohort IDs are recomputed for each device for each time period, where cohorts have a minimum size of $k=3$, effectively dividing the 6 devices into 2 cohorts (cohort IDs 1 and 2).
Consider the perspective of a third-party who observes the cohort IDs of devices on multiple sites. 
In any given period, they observe cohort IDs that are shared with a k-anonymous group of 3 devices.
Their goal, and the privacy risk to the users, is to identify unique cohort ID sequences in order to uniquely identify devices across the sites they visit. 

In the toy example, despite the minimum cohort size of $k=3$ for a single time period, after 2 time periods user devices 1 and 6 can be uniquely identified by their cohort ID sequences, [1,1] and [2,2], respectively. After 3 time periods, all users are uniquely identifiable by their cohort ID sequences.
When also using the fingerprinting data, user devices 3 and 4 are uniquely identifiable starting from period 1 (i.e. by [B, 1] and [B, 2]) and all user devices are uniquely identifiable by period 2 (i.e. by [A, 1, 1], [A, 1, 2], [B, 1, 2], [B, 2, 1], [C, 2, 1], [C, 2, 2]).

First-party cookies and partitioned storage provide at least two ways to track user cohort IDs, and more may arise in the future web ecosystem.
For example, third-parties embedded on a web site may store or access cohort ID sequences in partitioned storage, similarly to what is done with third-party cookies. 
Alternatively, first-party sites might store cohort ID sequences in first-party cookies and then share information with third-party trackers who have similar data sharing relationships with sites across the web. While this may sound convoluted, this kind of partnership between first-party sites and third-party trackers is similar to the current revenue generating model where first-party sites embed code from third-party trackers. The data collection model may change while the relationships remain similar.

\subsubsection{FLoC May Reveal Demographic Information About Users}
\label{section:concern floc may reveal demo information}

While k-anonymity can help protect users from being uniquely identified, cohorts are still subject to "homogeneity attacks" that can reveal sensitive information common to a cohort. Consider that cohorts are defined by browsing behavior, and some browsing behaviors are sensitive. An example of this risk provided by Google is where a cohort consists only of users who visited a website about a rare medical condition \cite{medinaMeasuringSensitivityCohorts}. In this case, by providing a user cohort ID, the FLoC API may reveal a user likely investigated that rare medical condition.

This risk also applies when considering demographic groups, such as those defined by racial background or household income. If certain demographics are highly prevalent in specific cohorts, then a FLoC cohort ID could reveal that a user is more likely in a certain demographic group. This privacy risk may have implications beyond simply leaking sensitive demographic information about supposedly anonymous users. For example, this may further expose users to online discrimination, such as price discrimination, predatory marketing, or targeted disinformation campaigns.

\subsection{Google's Approach To Sensitive Categories}
\label{section:google sensitive categories}

\begin{figure}
    \centering
    \includegraphics[width=0.45\textwidth]{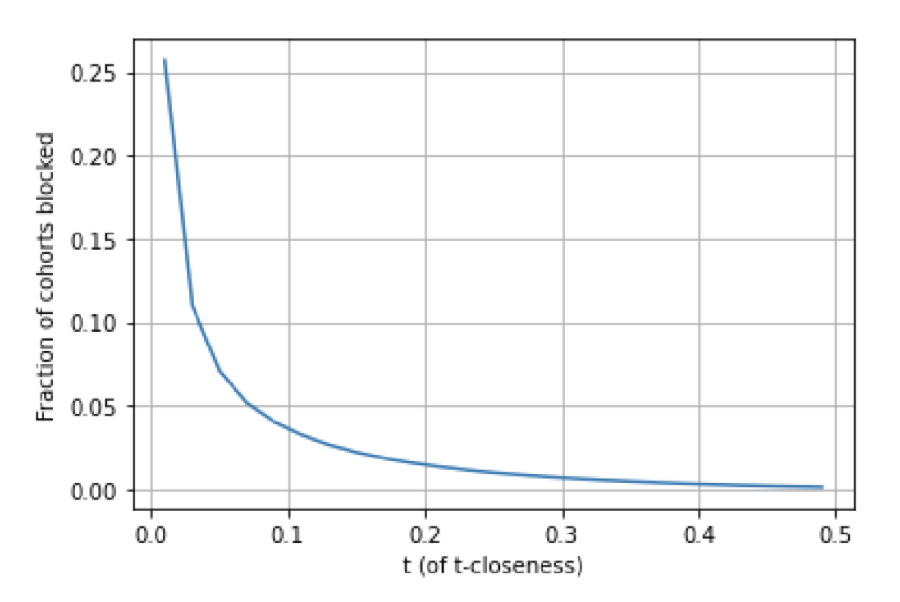}
    \caption{Plot provided by Google in \cite{medinaMeasuringSensitivityCohorts} showing the utility-privacy tradeoff between values of t for t-closeness and the fraction of cohorts that would then be blocked.}
    \label{fig:google_t_closeness_analysis}
\end{figure}

Google partly addressed the issue of sensitive categories in the implementation of FLoC that was deployed in the origin trial, but where the notion of sensitive categories was limited to sensitively categorized domains rather than user demographics~\cite{medinaMeasuringSensitivityCohorts}.

In particular, Google used the same sensitive interest categories defined for its
interest-based (personalized) advertising product~\cite{PersonalizedAdvertisingAdvertising}.  Google already forbids showing ads related to these categories or targeting a user based on them, which is why it also use these categories in this context.  It used these categories to determine whether a website is considered "sensitive" and then blocked cohorts that visited web pages related to a particular sensitive category at a much higher rate than the general population.  Examples of these categories include adult and medical websites as well as sites with political or religious content. 

Google noted that its methodology can be applied to other ontologies of sensitive categories \cite{medinaMeasuringSensitivityCohorts}. Here we describe Google's methodology which we then apply to sensitive categories defined by user demographics, namely racial background and household income.
To formalize the approach to blocking cohorts based on sensitive attributes, Google used t-closeness \cite{liTclosenessPrivacyKanonymity2007}. Cohorts satisfy t-closeness if for every sensitive category, the distribution of users who visited a web page related to that category has distance at most $t$ from the general distribution. We describe their approach to determining whether cohorts satisfy t-closeness as follows.

Assume each domain $d$ visited by a cohort $C$ has an associated category $X$. Cohorts can then be represented as a visit frequency distribution over categories. An \emph{anomalous} category, $X^{*}_C$ , is the category for a cohort $C$ that differs the most from the population frequency:

\begin{equation}
\label{eq:XC}
X^{*}_C = \arg\max_{X} CohortFreq(X, C) - PopulationFreq(X)
\end{equation}

Cohort $C$ does not satisfy t-closeness and is then deemed sensitive if the visit frequency for $X^{*}_C$ is greater than some threshold $t$ above the population frequency: 

\begin{equation}
\label{eq:t_closeness}
CohortFreq(X^{*}_C, C) - PopulationFreq(X) > t
\end{equation}

The origin trial adopted a $t=0.1$, blocking any cohort whose visit frequency to anomalous category $X^{*}_C$  appeared over 10\% more than the base population frequency. (The FLoC API returns no ID for blocked cohorts.) This resulted in 2.3\% of all cohorts being blocked in the origin trial \cite{FLoCOriginTrial}.

There is a clear utility-privacy tradeoff in determining the value of $t$: Lower thresholds of $t$ provide stronger privacy guarantees, but result in more cohorts being blocked, which degrades utility. 
To determine the value of $t = 0.1$ used in the origin trial, Google did preliminary analysis, using synced Chrome history data, to estimate the fraction of cohorts that would be blocked using various threshold values for $t$~\cite{medinaMeasuringSensitivityCohorts}. This is shown in Figure~\ref{fig:google_t_closeness_analysis}.

In the following work, we apply this methodology when considering users' demographics as sensitive categories. We note that Google names race as a sensitive category that should not be used in advertising in the same documentation used to categorize sensitive sites for its t-closeness analysis \cite{PersonalizedAdvertisingAdvertising}. Our analysis considers user racial backgrounds, as well as household income groups, as sensitive categories.

\section{Data And Computing Cohorts}
\label{section:data and computing cohorts}

To empirically test the privacy risks of FLoC, we leverage a dataset of user browsing history from comScore Inc.~ \cite{ComscoreTrustedCurrency}, allowing us to compute hypothetical FLoC cohorts from real-world browsing data. We first create samples of the browsing history data, filtering out data that do not meet the criteria used in Google’s FLoC OT. We then use the PrefixLSH hash-based clustering algorithm developed by Google to compute the cohorts used in the following analyses.
Further details about the data and analysis code can be found in our open source repository:
\url{https://github.com/aberke/floc-analysis}. 

\subsection{Data}
\label{section:data}

We use browsing sessions data from the comScore Web Behavior Database. We also use 2017 census data from the U.S. Census Bureau American Community Survey (ACS)~\cite{ACS:5YR}
and Current Population Survey (CPS) \cite{bureauCurrentPopulationSurvey}
in order to evaluate how well the browsing data represents the U.S. population as well as to create stratified samples that represent the U.S. population for use in analysis. 

The comScore Web Behavior Database was collected from 50,000 Internet users who gave comScore explicit permission to confidentially capture their detailed browsing behavior at the domain level. 
They were recruited via the Internet, with incentives including cash prizes and free software. They provided their data by installing tracking software on their machines to capture their web browsing sessions.

The sessions dataset is from the 52 weeks of 2017. Each session is associated with a unique machine ID and data rows include this machine ID, top-level domain name, session ID, timestamp, and associated statistics like number of pages viewed and duration of site visit.
An example representing sampled rows is provided in Table~\ref{table:sessions_example}.


\begin{table*}[ht]
\small
\begin{center}
\caption{Example (fake) data rows representing the web browsing sessions data. Sessions are associated with machine IDs and statistics about the session, as well as machine user demographics, which are categorically coded.}
\label{table:sessions_example}
\begin{tabular}{llllllllllll}
\hline
\textbf{Machine ID} & \textbf{Session ID}              & \textbf{Duration} & \textbf{Domain} & \textbf{Pages} & \textbf{Date} & \textbf{Time} & \textbf{…} & \textbf{Household Income} & \textbf{Race} & \textbf{Zip} \\
\hline
169007206           & 19308896 & 33                & site.biz        & 2                     & 20170515      & 7:25:23       &                             & 14                        & 1                          & 36832             \\
\hline
169007206           & 27157206 & 5                 & example.com     & 1                     & 20170515      & 8:36:55       &                             & 14                        & 1                          & 36832             \\
\hline
170422065           & 67238569                         & 46                & google.com      & 3                     & 20170515      & 23:27:22      &                             & 16                        & 1                          & 80233          \\
\hline 
\end{tabular}
\end{center}
\end{table*}

Machine IDs are also associated with a zip code and user demographic information reported at the household level. Demographics were self-reported when a user enrolled a machine with comScore. Multiple machines can be associated with a single household; machine IDs are not linked by household in the dataset and our analyses treat machines as independent. 
Before preprocessing, the sessions dataset includes 93,808 machines.

The demographic data we use in our analyses are household income, race, and zip code. Our analyses that use race are limited by the race categories reported in the comScore dataset, in which households are labeled as "Black", "White", "Asian", or "Other". We treat "Other" as being inclusive of multi-racial households and households that identify as Latino or Hispanic.
When considering income demographics, we use household income groups defined as "less than \$25,000", "\$25,000 - \$75,000", "\$75,000 - 150,000", "\$150,000 or more".

The demographic and geographic distribution of machines in the comScore dataset closely match census population estimates for the same period (2017). 
When comparing the distribution of the comScore population by U.S. state there is a Pearson correlation of  0.988 (p=0.000). When comparing racial background and household income groups there are Pearson correlations of  0.979 (p = 0.021) and 0.971 (p=0.029), respectively.
See Figures \ref{fig:comscore_demographics_state_pop} and \ref{fig:comscore_demographics_income_race} in the Appendix for more details. 

While highly correlated, there are differences between the demographics in our comScore dataset and the census population estimates.
These differences may be due to sampling bias and how demographics are counted for machine versus household. 
In our t-closeness analyses that consider user demographics, we mitigate these issues by using stratified sampling methods to create data panels that are more representative of the U.S. population with respect to race and income. This is further described in Section \ref{section:floc sensitivity to race and income}.

\subsection{Data Preprocessing}
\label{section:data preprocessing}

We preprocess the browsing sessions data by first filtering out domains that are not valid eTLD+1s~\cite{FederatedLearningCohorts2021}. We then sort and group browsing sessions by machine ID and week to create a dataset of records which we refer to as machine-weeks, where each record is the set of unique domains visited by a given machine for a given week. 
Machine-week records with fewer than 7 unique domains visited in the week are dropped, following the strategy used in the FLoC OT. (See Figure~\ref{fig:dist_domains_per_week} in the Appendix for the distribution of the number of unique domains per machine per week.) 

The resulting preprocessed dataset contains 2,073,405 machine-week records. 
It is used in each of the analyses that are further described: for computing the unicity of cohort ID sequences over time, for comparing user browsing behaviors by demographic groups, and for constructing panels that are representative samples of the U.S. population which are used in t-closeness analyses.
In each case, these analyses use the preprocessed dataset separately.

\subsection{Computing Cohorts}
\label{section:computing cohorts}

To compute cohorts similar to those used the FLoC OT, we use an open-source implementation of SimHash which has been used to replicate the OT results and has been verified by Google engineers \cite{ohtsuFLoCSimulator2022, aguilarFlocFlocSimulator2021}. We use this to compute a SimHash value for each machine-week to then be used as input in the PrefixLSH sorting algorithm. We compute cohort IDs using our own implementation of PrefixLSH. 

We might have instead used precomputed cohorts that were used in the FLoC OT by copying cohort mappings from Chrome browsers where the OT was implemented.
However, these cohorts definitions are not well-suited to our dataset, which is much smaller than what was used for the OT.
Directly using OT cohorts with our dataset would result in sparsely populated cohorts that are not $k$-anonymous.
Computing our own cohorts allows us to vary FLoC's $k$ parameter and the number of devices (machine-weeks) over which cohorts are computed, and better suit our dataset. It also allows us to manipulate these variables and investigate the relationship between dataset size, $k$, and cohort ID sequence unicity. Which values of $k$ we employ and how we use the resulting cohorts differs between our unicity and t-closeness analyses, described below.

\section{Unicity Of Cohort ID Sequences Across Time}
\label{section:unicity}

Section \ref{section:background} described the risk of FLoC enabling cross-site user tracking. In particular, third-parties may store or access users' cohort ID sequences across sites, and if these sequences are unique, they can be used as an identifier for third-parties to track users across the sites they visit.
This risk is also illustrated with a toy example in Table~\ref{table:cohort_id_sequences}.
In this section, we address this risk by analyzing the likelihood that a device has a unique cohort ID sequence after 4 periods, where cohorts are recomputed each period.
We note that while FLoC documentation describes how FLoC IDs are meant to be recomputed periodically, using browsing history from the previous 7 days, the frequency is unspecified.
In our analysis we use periods of 1 week (i.e. 7 days) and compute cohort IDs separately for each week, with no overlap in browsing histories between weeks.
Given we must make assumptions about the FLoC implementation, we aim to \emph{underestimate} unicity risks with the methods described below.

\subsection{Methods}

One issue we address here is dataset size.
We estimate the FLoC OT was implemented with more than N=100,000,000 devices, which was just a fraction of Chrome users.
This is orders of magnitude larger than the comScore dataset we use.
While we are interested in re-creating FLoC's OT as closely as possible, running FLoC with the same $k$ as the OT ($k=2000$) and our dataset's smaller $N$ creates far fewer cohorts than in the OT.
Fewer cohorts reduces the likelihood of devices having unique cohort ID sequences.
To address this limitation, we employ the following methods to expand our dataset size. We further use this dataset to investigate the relationship between $N$, the minimum cohort size $k$, and the risk of unicity. 

We effectively expand our dataset into one that represents a larger sample of 4-week sequences of browsing data, which we then use for analysis.
To do this, we split each machine's browsing data into sequences of 4 weeks. 
We then relabel the weeks associated with each machine-week by where they appear in a 4-week sequence, $\{1, 2, 3, 4\}$. 
We drop any sequence where the machine does not have sufficient data for all 4 weeks (7 unique domains per the FLoC algorithm). 
This process creates up to 13 4-week sequences (52 weeks / 4 weeks) of browsing data for each machine. 
This results in $N=305,312$ 4-week sequences, each representing a sample of a machine's browsing data over a 4-week period.
Cohorts are then computed for each of those 4 weeks across all machine-weeks represented in that week. Each 4-week sample then has a corresponding sequence of cohort IDs, $\{c_1, c_2, c_3, c_4\}$.


The sample cohort ID sequences in our dataset are more likely to be correlated than if the 4-week samples all came from different machines. This is because the machines contributing multiple 4-week samples to the dataset may have consistent browsing patterns across time.
This means our computed unicity risks are likely underestimates, as the likelihood of a sample's cohort ID sequence being unique is lower than in a real-world setting.

We also analyze how the risk of unicity increases when FLoC cohort IDs are combined with browser fingerprinting data. 
To do this we associate each 4-week sample with its machine's U.S. state 
and use the U.S. state as a weak proxy for fingerprinting. 
We use U.S. state because zip code is the only additional device-specific data we have for machines, yet it is too unique: many machines have unique zip codes in the comScore dataset. 
We aggregate zip codes to U.S. states.
Using this \emph{weak} proxy for browser fingerprinting data helps us measure how fingerprinting data might interact with FLoC and using a weak proxy is consistent with our goal to produce conservative underestimates of unicity risks.

To estimate the risk of unicity over time, we then count the number of samples in our dataset with unique cohort ID sequences after each of the 4 weeks, as well as how many such sequences are unique when combined with their U.S. state.

\subsection{Findings}
\label{section:unicity findings}

\begin{figure*}[h!t]
     \centering
     \begin{subfigure}[b]{0.32\textwidth}
         \centering
     \includegraphics[width=\textwidth]{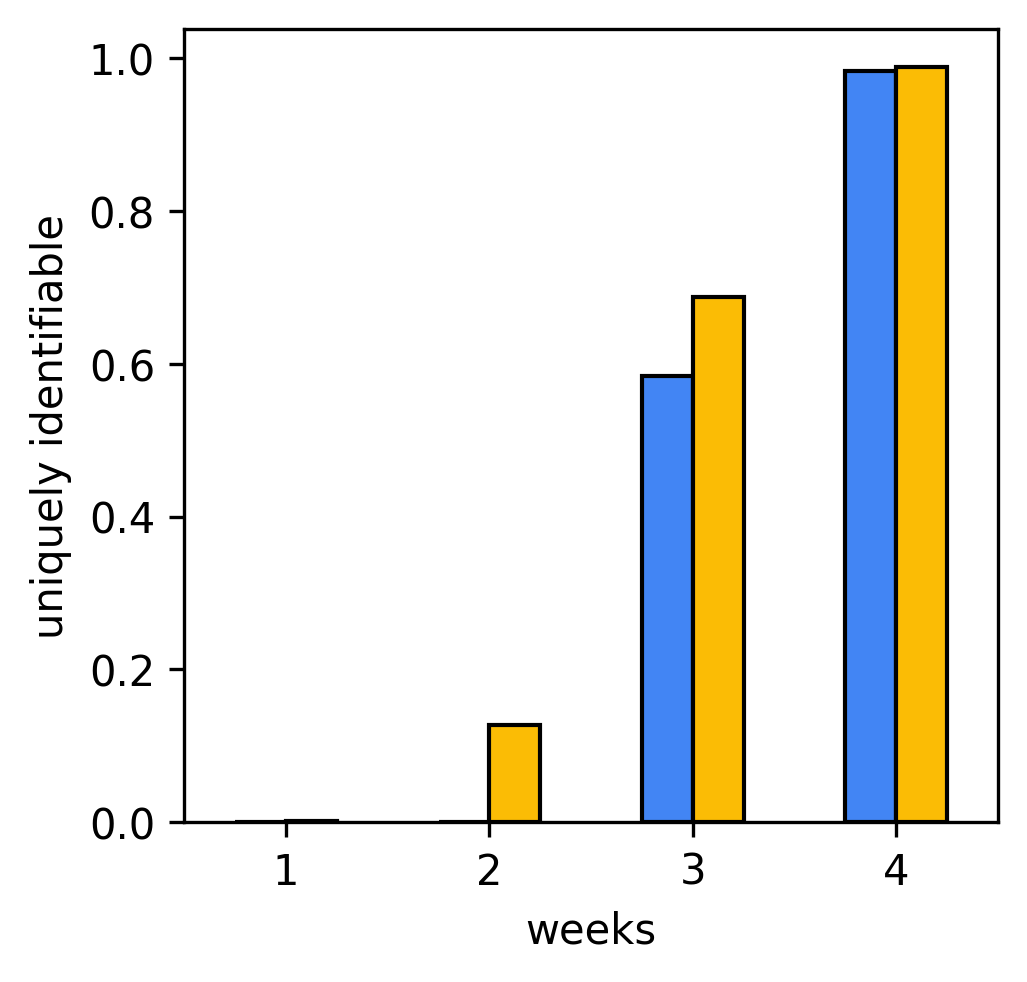}
         \caption{}
     \end{subfigure}
     \hfill
     \begin{subfigure}[b]{0.33\textwidth}
         \centering
         \includegraphics[width=\textwidth]{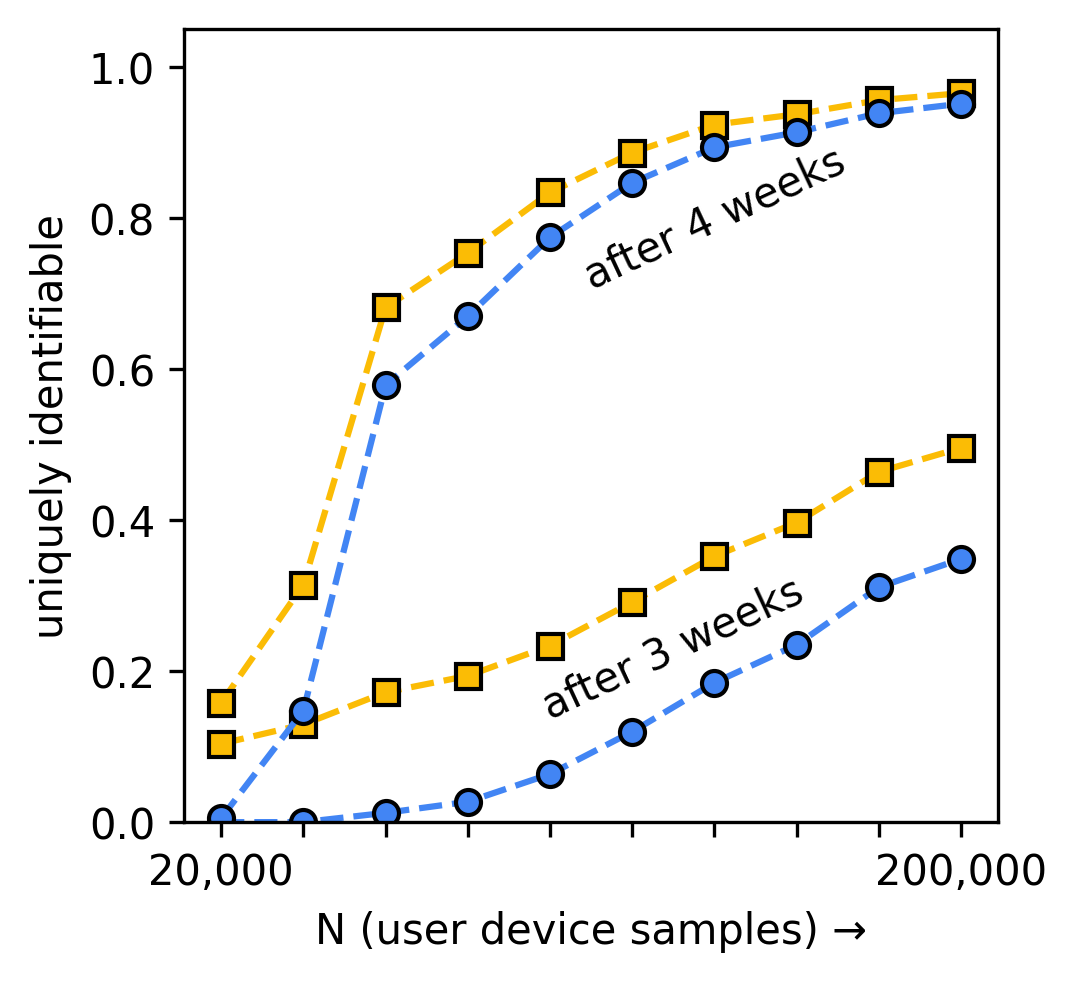}
         \caption{}
     \end{subfigure}
     \hfill
     \begin{subfigure}[b]{0.32\textwidth}
         \centering
         \includegraphics[width=\textwidth]{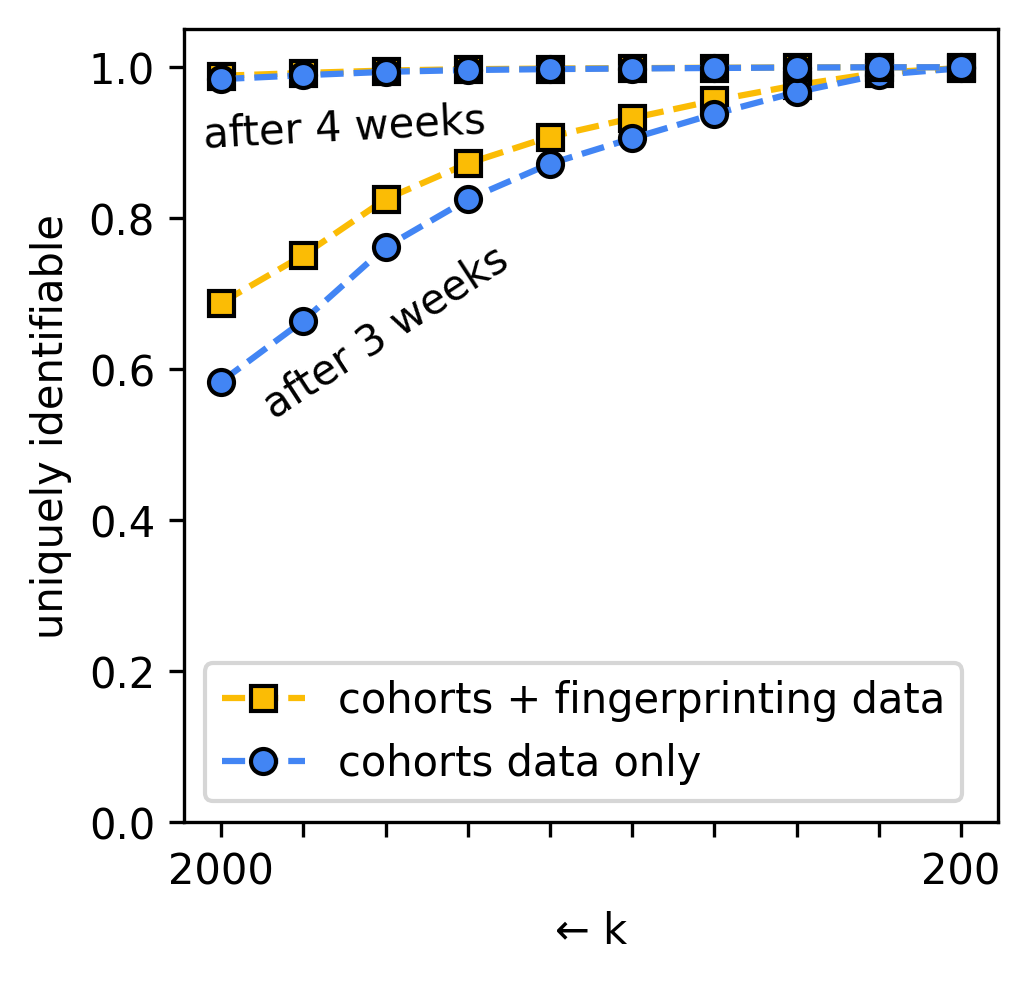}
         \caption{}
     \end{subfigure}
        \caption{(a) Unicity of cohort ID sequences, across weeks. Computed using k=2000 and N=305,312 4-week sequences of data sampled from user devices. Blue bars indicate the fraction uniquely identifiable by cohort ID sequence data alone, after 1, 2, 3, 4 weeks of observation. Orange bars indicate the estimated enhanced risk when combining cohort ID data with available fingerprinting data. The residential U.S. state for the device user is used as a weak proxy for fingerprinting data. The dataset used represents a small fraction of Chrome and potential FLoC users; reported results are underestimates. More user devices for a fixed k results in more cohorts and higher likelihood of unique cohort ID sequences.
(b) Fraction of uniquely identifiable samples after 3 and 4 weeks, for a fixed k=2000 and varying N. More devices result in more cohorts, increasing the risk of unique cohort ID sequences over time.
(c) Fraction of uniquely identifiable samples after 3 and 4 weeks, using fixed sample size N=305,312 and diminishing values of k. For a fixed N, a lower k results in more cohorts.}
\label{fig:unicity_results}
\end{figure*}

We find that more than 50\% of samples from user devices are uniquely identifiable after 3 weeks, and that this risk climbs to 95\% of samples after 4 weeks.
These risks are even higher when cohort IDs are combined with our weak fingerprinting proxy, U.S. state, which provides another dimension for unicity. See Figure~\ref{fig:unicity_results}~(a). As a point of comparison, a 2018 analysis found that 33\% of devices can be uniquely identified using fingerprinting alone, and projected this risk was in decline due to trends in browser developments  \cite{gomez-boixHidingCrowdAnalysis2018}.

We consider our risk estimates conservative underestimates given our dataset size.
Our analysis used $N$=305,312 samples, each representing 4-weeks of user browsing data, and cohorts with size $k=2000$ were computed for each week in the 4 weeks, resulting in 97-100 cohorts (varying by week). This sample size and the resulting number of cohorts are both small when considering the number of potential FLoC users. For example, while the  FLoC OT included only  a small fraction of Chrome users, it resulted in 33,872 cohorts with a minimum cohort size of $k=2000$.

The risk of unique cohort ID sequences increases with more cohorts. When $k$ is kept constant, more user devices ($N$) result in more cohorts, and hence more potentially unique sequences of cohort IDs from week to week.
We test how the number of devices impacts unicity by fixing $k=2000$ and varying $N$ from $N=$20,000 to $N=$200,000 in increments of 20,000. 
For each value of $N$, we randomly sample $N$ 4-week sequences from our dataset and compute corresponding cohort IDs.
Figure~\ref{fig:unicity_results}~(b) shows that the fraction of these sequences that are unique after 3 and 4 weeks increases as $N$ increases. 
We note this result that more user devices results in higher risk of unicity is counterintuitive at first glance; we remind the reader to consider that the FLoC OT clustering algorithm effectively limits the average size of cohorts for a fixed k, where here $k=2000$. 
Overall, more user devices for a fixed $k$ results in more cohorts, increasing the possible combinations of cohort IDs and therefore likelihood of unique cohort ID sequences.

We remind the reader that $k$ is a tunable parameter in the FLoC algorithm, where $k$ determines the minimum cohort size in order to provide $k$-anonymity.
We also explore how $k$  impacts unicity for a fixed $N$. Figure~\ref{fig:unicity_results}~(c) shows the fraction of uniquely identifiable samples from user devices after 3 and 4 weeks, using a fixed N=305,312 and values of $k$ ranging from 2000 to 200, diminishing in increments of 200. As $k$ decreases, there is a steady increase in the fraction of uniquely identifiable sequences by week 3. Higher values of $k$ lead to fewer cohorts with more users in each cohort—and hence higher levels of privacy.

\section{FloC’s Sensitivity To Race and Income}
\label{section:floc sensitivity to race and income}

This section addresses the concern described in Section \ref{section:concern floc may reveal demo information} that FLoC may leak sensitive demographic information about users. 
Specifically, we assess this risk for user race and household income. 
We use the t-closeness methodology developed by Google in their handling of sensitive categories of websites, as described in Section~\ref{section:google sensitive categories}.
We note that this application of t-closeness may not fully measure the risk of FLoC leaking sensitive demographic information, or otherwise contributing to online discrimination. However, our goal is to assess an important risk that Google did not address by applying Google's own methodology to an ontology of sensitive categories that Google did not report on. 

Before our t-closeness analyses, we first demonstrate that there are significant differences in browsing behaviors by demographic group in our dataset. This is important because if different demographic groups in our dataset exhibited no differences in browsing history, we should expect a clustering algorithm, such as FLoC, to cluster users independently of demographics.

\subsection{Browsing Behavior Differences By Demographic Group}
\label{section:browsing behavior differences}

\subsubsection{Methods}
\label{section:browsing behavior differences methods}

To test whether different demographic groups exhibit significant differences in browsing behavior, we compare domain visit frequencies across demographic groups using chi-squared tests of independence.
We use the dataset of machine-weeks described in Section~\ref{section:data preprocessing}, where each record is the set of unique domains visited by a given machine for a given week. We count a domain visit as a visit to a unique domain within a single week (so repeat visits to the same domain within one week are not counted), similar to FLoC.

We test race and income separately.
In each case, we divide the machines in our dataset into subpopulations defined by their demographic groups.
We use the categories described in Section~\ref{section:data}. Our subpopulations for race are machines labeled as  White, Black, Asian, and Other. For income groups, we use "less than \$25,000", "\$25,000 - \$75,000", "\$75,000 - 150,000", and "\$150,000 or more". 
We create and test an additional random control group, randomly drawn  from the aggregate population without replacement ($n = 0.25 \times N$) as a robustness check and point of comparison.

For both race and income, we run a series of chi-square tests of independence with the top $D$ domains in our dataset, progressively increasing $D$ by 10 so that $D = {10, 20, 30, 40, \dots, 100}$. For each value of $D$, the test compares a subpopulation’s distribution of visit frequences of the top $D$ domains to the distribution of the aggregate population. In particular, we test the following null hypothesis separately for each subpopulation with $p=0.01$: the frequency of visits to the top $D$ domains by the subpopulation matches the frequency of visits by the aggregate population.

\subsubsection{Findings}
\label{section:browsing behavior differences findings}

\begin{figure*}[ht]
    \centering
    \includegraphics[width=\textwidth]{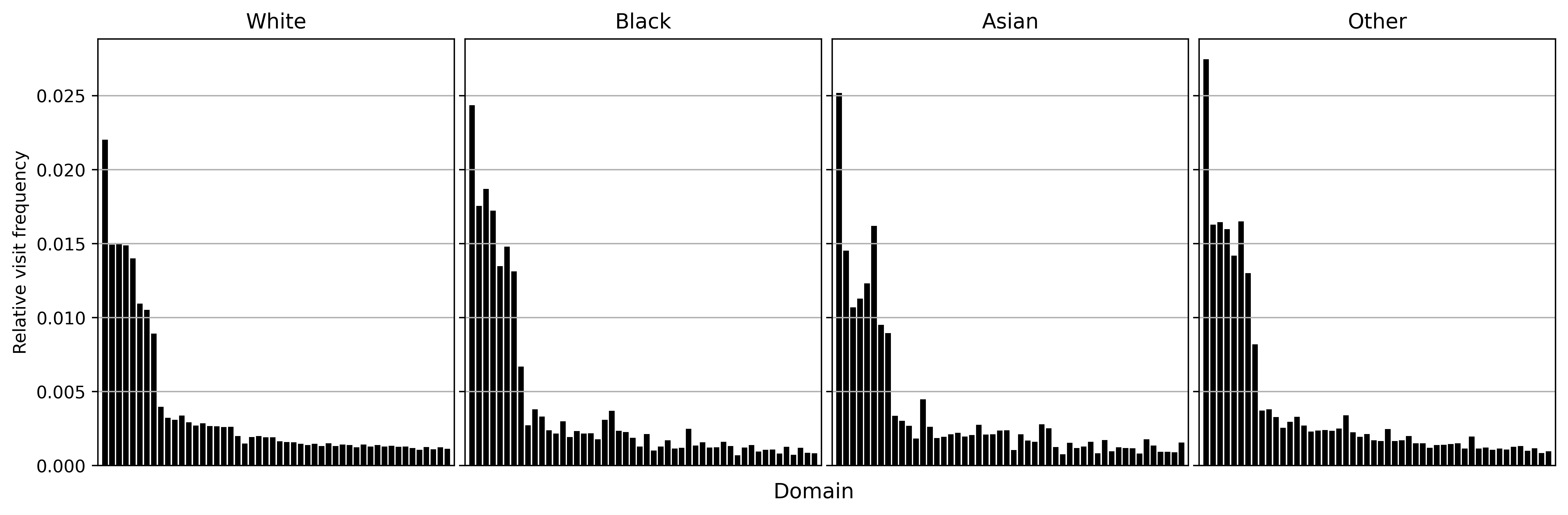}
    \caption{Relative visit frequencies for the top D=50 domains, compared by racial background. Domains are sorted by visit frequency in the overall population. (See Section~\ref{appendix:domains frequencies} in the Appendix for domain names and visit frequency values.)}
    \label{fig:top50_domains}
\end{figure*}

We find that different demographic groups exhibit significant differences in browsing behavior. 
For each of the chi-square tests of independence described in Section~\ref{section:browsing behavior differences methods}, we find a statistically significant difference in domain visit frequency ($p < 0.0001$) from the aggregate domain visit distribution. 
This is the case for each set of $D$ top domains, with $D$ ranging from 10 to 100, for each subpopulation defined by race or income groups.
We note the p-values are small enough that a Bonferroni correction for our multiple tests has no impact on determining significance.
In contrast, our randomly drawn subpopulation shows no significant difference (with $p > 0.75$) for each $D$. 
These results indicate that each subpopulation, defined by race or income, varies significantly from average browsing behavior. 

The differences in domain visitation frequency between racial groups are illustrated in Figure~\ref{fig:top50_domains}, which shows the frequency of visits to the top $D=50$ domains. 
See Section~\ref{appendix:domains frequencies} in the Appendix for domain names and relative visit frequencies for the aggregate population and each subpopulation.
While the overall shape of each group’s distribution is similar, large differences in the relative frequencies and order of top domains can be seen between the groups.  We see similar results for the subpopulations defined by income, but omit such a plot for brevity. 

\subsection{t-closeness By Race and Income}

\subsubsection{Methods}
\label{section:t-closeness methods}

\begin{figure*}[ht]
    \centering
    \includegraphics[width=\textwidth]{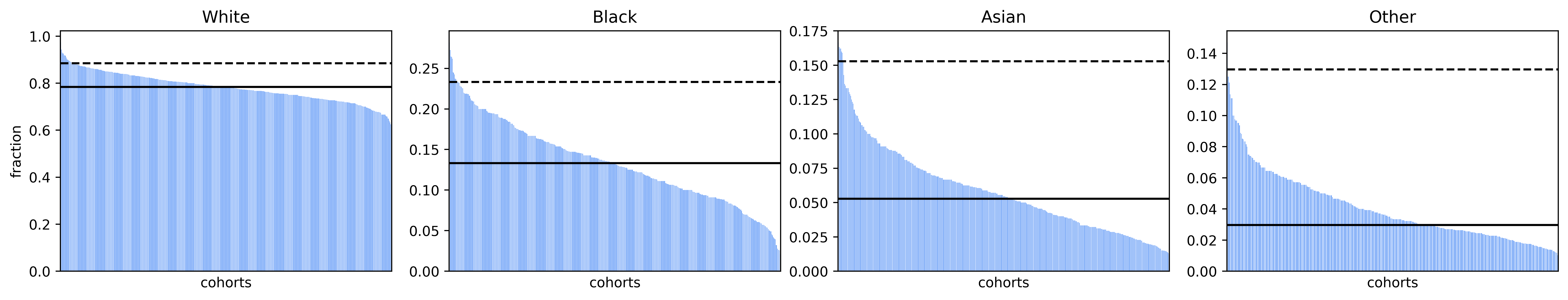}
    \caption{Demonstration of computing t-closeness across racial categories for a single panel. Each of the 4 plots illustrates analysis over the same set of cohorts computed using the FloC algorithm for this single panel, but where each plot evaluates the prevalence of a different racial category across these cohorts. In each plot, each vertical blue bar represents a single cohort. The bar height corresponds to the fraction of cohort members that are of the given racial category. Cohorts are ordered from most to least prevalent in each plot. Horizontal solid lines represent the mean prevalence for that category in the overall population. Dashed lines represent the threshold above which t-closeness for $t=0.1$ would be violated. While many cohorts over-represent various racial categories to varying degrees, very few violate t-closeness at the $t=0.1$ level.}
    \label{fig:t_closeness_single_panel}
\end{figure*}

To calculate t-closeness based on users’ race and income demographics, we follow the same definitions used by Google in~\cite{medinaMeasuringSensitivityCohorts}. These are described in Section \ref{section:google sensitive categories}, and formally defined by equations \ref{eq:XC} and \ref{eq:t_closeness}.
In Google's analysis, $X$ represents a website category visited by users in a cohort.
Here, $X$ represents a demographic category for users in a cohort (i.e. race or income group).

To better match Google's definition, we refer to the number of user machines of a given demographic group as frequency.
We compute how many cohorts violate t-closeness by comparing the relative frequency of each demographic group in each cohort to their relative frequency in the overall population. 
A cohort $C$ is considered sensitive if there is a demographic group $X^{*}_C$ that differs by at least $t$ from the overall population frequency: 
\[CohortFreq(X^{*}_C, C) - PopulationFreq(X) > t
\] 
We compute t-closeness for the income and race demographic groups separately.

We create our empirical dataset to measure t-closeness by sampling panels from the comScore data that are representative of the U.S. population with respect to race and income.
We do this so that results using our limited dataset can then be extrapolated to the larger U.S. population.

We use the preprocessed machine-weeks dataset described in Section \ref{section:data preprocessing}.
For each week in our dataset, we create 10 representative panels. 
To ensure that each panel has a demographic distribution matching the broader U.S. population, we use stratified random sampling.
To do this, we use the joint distribution of household income and race for the overall U.S. population, as reported in the census 2017 CPS data, using the income and race groups described in Section~\ref{section:data}. We create each panel by sampling machines without replacement from the comScore dataset until each panel matches the joint distribution in the CPS data exactly. Each panel is then representative of the broader U.S. population with respect to both race and income, without double-counting machines in any panel.
Panel sizes vary by week, as the size of the underlying dataset of machine-weeks varies by week. Panels computed for the same week are of the same size. Across weeks, panels have a mean size of 21,875 with a standard deviation of 1,421.

Given each panel represents a user population's browsing history over a unique 7-day period, we treat each panel as a separate dataset when running the FLoC algorithm. Computing cohort IDs within each panel separately means we calculate cohort assignments for 520 different sampled populations (10 panels for each of 52 weeks). This allows us to report means with confidence intervals for the fraction of cohorts that violate t-closeness for a given $t$.

Figure \ref{fig:t_closeness_single_panel} demonstrates computing t-closeness metrics for a single panel across race. 
Each subplot illustrates analysis over the same set of cohorts computed using the FloC algorithm for this single panel, but where each plot evaluates the prevalence of a different racial category across these cohorts.
Each vertical bar represents a single cohort in the panel. For each race category, the height of a cohort's bar represents the fraction of it's members that are of that race. Cohorts are ordered from most to least prevalent for that race category. In each plot, a horizontal black solid line represents the mean prevalence for that category in the overall population, $PopulationFreq(X)$, while the dashed line represents the threshold where t-closeness for $t=0.1$ would be violated, $PopulationFreq(X) + t$. 
As an example, consider t-closeness applied to Black users. The black line in the second plot in Figure~\ref{fig:t_closeness_single_panel} shows that Black users make up 13\% of the overall panel population. Many cohorts over-represent Black users, but only a small group at the left edge of the plot would exceed the dashed line threshold and therefore violate t-closeness for $t=0.1$.

The FLoC OT used a minimum cohort size of $k=2000$, resulting in 33,872 cohorts.
There is a relationship between $k$ and the total number of cohorts that impacts both utility and privacy. 
Our panels are small relative to the FLoC OT,
 which necessitates using a smaller minimum cohort size $k$.
For example, using $k=2000$ with our dataset results in only several cohorts per panel, meaning that users who exhibit very different browsing behaviors are more likely grouped into the same cohort.
To choose $k$, we observe that in the FLoC OT, $\frac{|cohorts|}{k} = \frac{33872}{2000} \sim 16.9$. We empirically test which value of $k$ would most closely result in a similar ratio. The analysis presented below uses $k=30$, which closely follows this ratio. Because different values of $k$ might impact t-closeness results, we perform robustness tests where we run the same analyses for $k=25$, $k=35$, and $k=2000$. We find these various values of $k$ yield similar results.

Given that our panel and cohort sizes are relatively small, there is some probability that any given cohort will violate t-closeness purely by chance.
To test whether our empirical results show t-closeness violations beyond what is expected by chance, we compare our results with two baselines. The first baseline is created by randomly shuffling users' SimHash values after panel creation. This creates a randomized version of our panel data where cohort IDs are guaranteed independent from both race and income. 
For the second baseline, we model the expected number of cohorts violating t-closeness by using a binomial CDF. Given a mean cohort size $n$, and the marginal probability of a user being of race or income group $r$ as $p_r$, a threshold, $k_r$, for the number of users from race or income $r$ can be set to satisfy t-closeness for a given $t$ as $k_r = n \times (p_r + t)$. Then the expected fraction of cohorts that contain at least $k_r$ members of demographic $r$ can be modeled as $1 - F_r(k_r; n, pr)$ where  

\[
F_r(k_r; n, p_r) = Pr(X_r \leq k_r) = \sum_{i=0}^{\lfloor k_r \rfloor} {n \choose i} p_r^{i} (1 - p_r)^{n-i}
\]

We validate that this model accurately simulates the expected number of cohorts violating t-closeness due to random chance by comparing it to our randomized panel data, finding them effectively indistinguishable. 

We also then check whether t-closeness with respect to race or income would be violated at the scale of the OT in the case that cohorts were assigned independently of demographics. We do this by constructing a control panel where the size and number of cohorts matches the FLoC OT. We randomly assign race and income to panel users with a distribution that matches the U.S. population, based on the CPS 2017 census estimates. With this control dataset we find that t-closeness would not be violated with the value of $t=0.1$ used in the OT. Details are in the Appendix.

\subsubsection{Findings}
\label{section:t-closeness findings}

\begin{figure*}[ht]
    \centering
    \begin{subfigure}{\textwidth}
        \includegraphics[width=\textwidth]{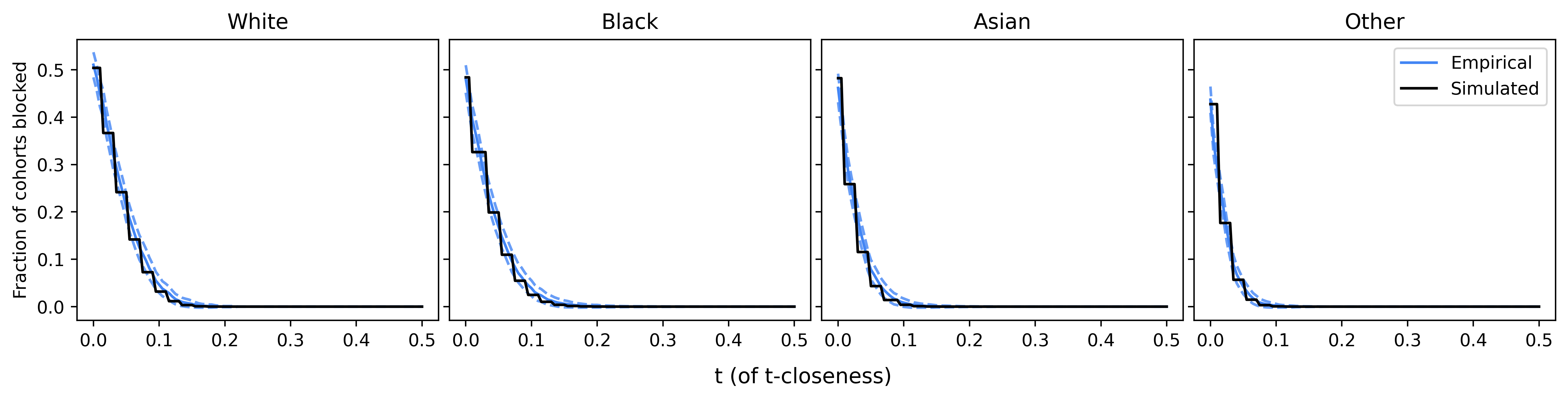}
         \caption{}
    \end{subfigure}
    \begin{subfigure}{\textwidth}
        \includegraphics[width=\textwidth]{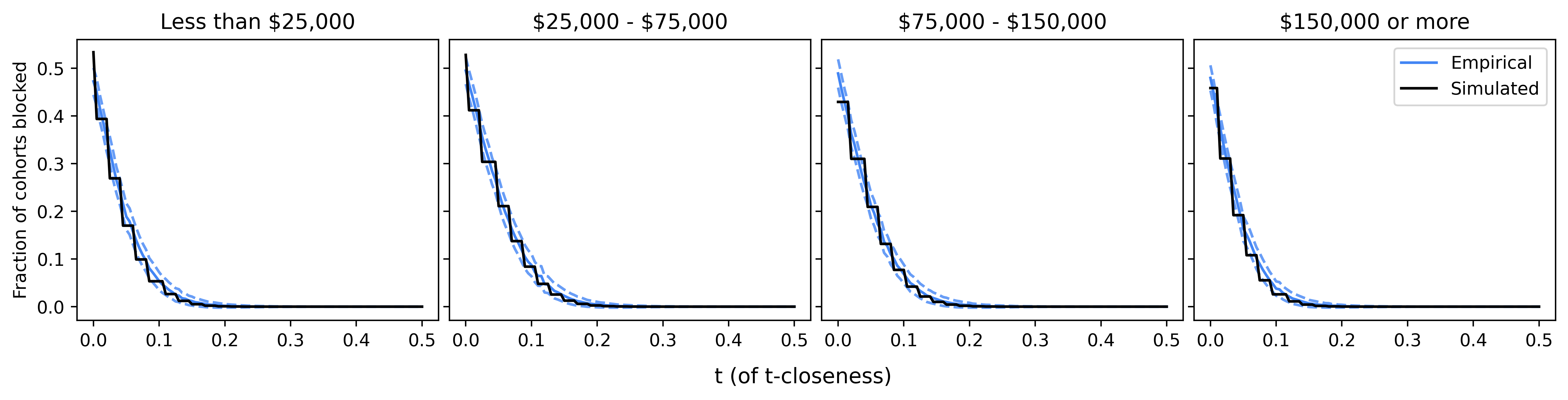}
         \caption{}
    \end{subfigure}
    \caption{FLoC t-closeness violations using race as a sensitive category, for varying levels of t. (Compare to Figure~\ref{fig:google_t_closeness_analysis} from Google's sensitivity of cohorts analysis~\cite{medinaMeasuringSensitivityCohorts}.) Solid blue lines indicate mean empirical values calculated over 520 panels, dashed lines indicate 95\% CIs. Black lines indicate t-closeness violations expected by random chance, simulated by a binomial CDF. The simulated values fall within the empirical results 95\% CIs.}
    \label{fig:t_closeness_by_race_income}
\end{figure*}

We do not find that the likelihood of a cohort violating t-closeness with respect to race or income is any greater than random chance, when using the FLoC OT algorithm and our dataset.
Figure~\ref{fig:t_closeness_by_race_income} shows t-closeness curves for each race and income group versus a simulated baseline that represents the likelihood of violating t-closeness by random chance. 
The simulated baseline was generated using a binomial CDF, described in Section \ref{section:t-closeness methods}. We find no distinguishable difference between our empirical data and this baseline for either income or race — the simulated values that represent t-closeness violations due to random chance are almost exclusively within the empirical data’s 95\% confidence interval. 
We also compared our empirical results to a random baseline generated by shuffling cohort IDs, described in Section~\ref{section:t-closeness methods}, with the same result.
For robustness tests, we ran these same analyses with values of $k=25,30,35,2000$, each time finding that the simulated baselines fall within the 95\% CIs for the empirical data.

In Section \ref{section:t-closeness methods}, we also show how a control panel created with a similar number of users and cohorts as the FLoC OT, and where cohort IDs were randomly distributed independent of demographics, would not violate t-closeness. If results from our limited dataset are considered a good proxy for larger datasets, this might imply that t-closeness would not have been violated with respect to race or income in the OT, or a similarly sized deployment.

These t-closeness results are surprising, given the significant differences in the frequency of domain visits between demographic groups we observe in our dataset, as described in Section \ref{section:browsing behavior differences}. We consider potential explanations for this, and their implications, in the discussion. 

\section{Discussion}

Following the completion of the FLoC trial, Google published simple metrics from the trial about the total number of cohorts and the number that were dropped due to their sensitivity definition~\cite{FLoCOriginTrial}. However, Google did not report on the success of FLoC in terms of clustering or the potential utility for interest-based advertising. 
The PrefixLSH algorithm used in the FLoC OT and this work was described as an initial implementation, subject to change.
Their previous analysis evaluated this algorithm by using datasets that were not browsing histories, and how well this algorithm would perform within a browsing platform was left unclear and noted as a limitation of their work \cite{epastoClusteringPrivateInterestbased2021}.

Our analyses show clear privacy risks and raises questions about the utility and privacy tradeoffs of the FLoC algorithm, which may have contributed to Google canceling its proposed implementation.

\subsection{FLoC Enables Individualized Tracking}

Although FLoC was designed with the goal of mitigating the risks of individualized tracking of users across the web, it does not achieve this goal.
We find that FLoC cohort IDs enable the unique identification, and hence individualized tracking of user devices. This is the case for more than 50\% of user devices in our dataset after only 3 weeks, and this risk climbs to more than 95\% after 4 weeks. We also show how when cohort data is combined with other fingerprinting mechanisms, these risks increase.
Our results present underestimates of these risks due to the limited size of our dataset, which we show by plotting the relationship between these identification risks and dataset size. 

The advertising and publishing industry is a dense network of actors that already share an enormous amount of information about browser users. 
Section \ref{section:background} described how partitioned storage and first-party cookies can be leveraged to identify and track users by their FLoC cohort ID sequences.
If tracking FLoC cohort IDs can aid in the kind of tracking that makes the current ad-serving ecosystem profitable, it is reasonable to assume that these actors will do it.

It's important to recognize that the identification risk posed by FLoC cohort IDs behaves differently than the risks of browser fingerprinting that are more commonly studied.
Overall, prior work has found that browser fingerprinting techniques uniquely identify roughly 33\% of devices ~\cite{gomez-boixHidingCrowdAnalysis2018}.
In practice, the real-world risk decreases as the number of devices observed by a third-party increases: a device is more likely to be unique in a pool of 20 devices through fingerprinting than a pool of 200. 
FLoC cohort sequences behave differently. 
Because $k$ modulates the minimum size of cohorts, the number of cohorts increases with the number of devices. 
As explained in Section ~\ref{section:unicity findings} and shown in Figure ~\ref{fig:unicity_results}~(b), this increases unicity risk. This means FLoC is \textit{more} useful for identifying users as the number of users increases, in contrast to common fingerprinting methods.

We consider a few potential strategies to mitigate this privacy risk of individualized tracking, but each has negative impacts in terms of utility. A simple strategy would be to avoid updating user cohort IDs. However, FLoC would then provide a less relevant signal for advertisers.
Another potential strategy is to frequently clear all browser data (e.g. first-party cookies), to prevent the accumulation of cohort ID sequences over time. 
However, this would negatively impact usability for browser users, as browser data and first-party cookies enable features, such as keeping users logged in or maintaining consistent e-commerce shopping carts. If this tactic were pursued, the frequency of clearing the data would present a precarious privacy-utility tradeoff.
In addition, cohort sizes could be increased, providing greater k-anonymity guarantees. Such a strategy might be further motivated by the clear relationship between the minimum cohort size ($k$) and the likelihood of uniquely identifying users over time, shown in our empirical results. This strategy might even be combined with the strategy of clearing browser data, to allow clearing data less frequently. However, increasing cohort sizes would make FLoC less effective at separating users by more specific interests, and would again make cohort IDs less useful for interest-based advertising.

\subsection{Browsing Behaviors Significantly Differ By Demographics, But Cohorts Do Not}

A major criticism of FLoC was that it may leak sensitive demographic information about browser users. Google addresses a related issue by computing a specific definition of cohort sensitivity, defined by the frequency with which a cohort visited domains that Google considered ``sensitive". Although Google recognizes that race is another sensitive attribute that should not be used by advertisers, Google did not address sensitivity based on cohort members' race or other demographics.

We address this risk. First, we show that browsing behaviors significantly differ by race and income.
In particular, we show that the browsing behaviors of different demographic groups, defined by race and income, significantly differ from the overall population in our dataset when counting the frequency of unique domains visits by week, similar to the FLoC algorithm. 
One might then expect a clustering algorithm based on these data to create some clusters (cohorts) grouped around certain user demographics. 
In such a scenario, users' cohorts could then aid in the inference of their race or income.

However, we did not find that the risk of cohorts leaking information about users' race or income, when using the FLoC OT algorithm, was any greater than if cohorts were randomly assigned.
We note that these findings are based on t-closeness analyses, using methodology proposed by Google. 
It is possible that other methods may better find relationships between user demographics and FLoC OT cohorts.
Furthermore, our analyses are limited to the specifics of our dataset. Our dataset and resulting cohorts were small and actual analysis of the FLoC OT data or a larger real-world deployment might yield different results. Also, the comScore dataset was collected from Internet users who signed up to have their browsing history tracked for cash prizes and other incentives.
It's possible that such a selection bias could cause the browsing behavior of users in our dataset to overall be more similar versus the larger U.S. population. The fact that we find significant differences in browsing behavior by demographic groups suggests that this is not the case, but it is not conclusive.

Regardless, we highlight three potential explanations for our seemingly contrasting results: 
(1) The differences between the browsing patterns of different race and income groups in our data may not be large enough to cause cohorts to overly represent these groups, or 
(2) FLoC is able to effectively cluster based on browsing history on dimensions independent of race and income, or (3) FLoC does not effectively cluster based on browsing history.

Despite which explanation may be most plausible, the FLoC OT algorithm was considered a preliminary implementation and it is likely that improvements to the algorithm, or alternative web user clustering strategies, could heighten the risk of leaking information about users' demographics. This presents a tradeoff between utility and privacy that could be avoided without FLoC.

\subsection{The Privacy-Utility Tradeoff of Interest-Based Advertising}

The conflict between user privacy and advertiser utility has been explored by researchers, artists, and technology designers for over a decade \cite{kazienko_adrosaadaptive_2007,juels_targeted_2001,guha_privad_2011,toubiana_adnostic_2010,kontaxis_tracking_2015,howe_engineering_2017}. Many of these projects adopt a similar approach to FLoC, where "interest profiles" are calculated in-browser. However, in projects such as \cite{toubiana_adnostic_2010} and \cite{juels_targeted_2001}, these profiles are either not made available to ad networks, or the advertising selection is done entirely client-side from a "cache" of available advertisements. Both of these approaches, if implemented, could greatly reduce the amount of identifiable information collected by advertisers.  Other projects take a more adversarial approach, testing and deploying technologies that obfuscate user identities and frustrate online advertisers directly \cite{howe_engineering_2017}. 

However, recent research calls into question the extent of the privacy-utility tradeoff these works explore. The overall case for interest-based advertising is that it increases revenue for both advertisers and ad publishers. Yet recent studies estimate that selling targeted ads increases ad publishers' revenue by only 4\% \cite{marotta_online_2019, marotta_welfare_2022}. This suggests that behavioral targeting, while it might increase advertising campaign effectiveness, does not benefit all players. Is another approach feasible?

\subsection{Recommendation: Contextual Ads Without Individualized Tracking }

Interest-based advertising is just one mechanism to reach users with relevant ads. Another is first-party contextual information. With such a paradigm, relevant ads for a user are based on the content of the page the user is visiting, rather than their past behaviors. For years, this was the dominant ad-serving paradigm, providing a clear example for alternative models that do not necessitate individualized user tracking. 

This approach may not negatively impact web generated revenue as much as it first appears. After the implementation of the GDPR, \textit{The New York Times} ended programmatic ad buying in Europe entirely. This forced ad buyers to purchase advertising space on their site without any personalized targeting. As noted in \cite{hwangSubprimeAttentionCrisis2020}, this should have in theory reduced their revenue—advertisers should have ostensibly purchased more "effective" ad space in spots that offered targeting. Yet, \textit{The Times}' advertising revenue didn't change~\cite{davies_after_2019}.

Context-focused advertising alternatives could take many forms with respect to how third-parties or browsers learn what ads could be relevant based on a page’s content or their audiences. 
While such a shift would require an overhaul of the ad-serving ecosystem, content-based advertising strategies could sidestep the kinds of privacy risks raised in this paper. 

FLoC was a first step in this direction, yet it was still premised on tracking users and associating them with identifiers. Our unicity results demonstrated the difficulty of achieving privacy with such a system premise - even if an identifier represents a larger group (cohort), it provides one more vector for uniquely identifying devices.

Google canceled FLoC in favor of a new approach (Topics), which is a more privacy-preserving step in the direction towards advertising based on contextual content. However, like FLoC, it is still premised on tracking users' browsing behaviors and future work will need to demonstrate that this new approach goes far enough to preserve user privacy.

\subsection{Recommendations For Future Systems}

As developers iterate on solutions to the online advertising ecosystem and move past third-party cookies, sensitivity and risk assessments should be incorporated into the development of their system designs.

While proposals such as FLoC often integrate a public feedback phase for comment, formal tests of these new systems are scarce and difficult to execute. Our analysis required us to re-implement FloC based on design documents and notes from FLoC engineers and to leverage a proprietary dataset of browsing histories. These hurdles make analysis by and for the public inaccessible to many external researchers and community members. 

We propose that future designers of such systems incorporate a broader array of sensitivity analyses that include demographic information. We also call on developers to publish tools, example datasets, and code so that their proposals can be more easily interrogated by researchers.

\section{Conclusion}

In this paper, we provide a post-mortem empirical analysis of two major privacy concerns raised about FLoC, which Google designed and tested as a means to facilitate interest-based advertising on the web without third-party cookies. First, contrary to its core aims, FLoC enables the tracking of individual users across sites.
We find that more than 95\% of cohort ID sequences sampled from devices in our dataset are uniquely identifiable after only 4 weeks.
We then show how our results represent underestimates of this risk, and that risk increases with the use of common device fingerprinting methods. 
We also show that there is a relationship between sensitive user demographics (namely race and income) and browsing behaviors. Yet despite this relationship, we did not find that the FLoC algorithm tested by Google meaningfully groups users by race or income. 
Our analysis highlights risks that should be considered in future proposed solutions for interest-based advertising on the web, and demonstrates strategies for other researchers seeking to interrogate the privacy implications of such proposals.


\begin{acks}
The authors thank Latanya Sweeney for providing guidance and resources for this project, and Bennett Cyphers at the EFF for helpful public analysis of FLoC. Thanks also to Eric Rescorla and Martin Thomson at Mozilla for their work that inspired our analysis and critique, and Martin’s helpful comments on an early draft of this article. 
Thanks to anonymous reviewers, especially reviewer “B", whose critical and constructive comments greatly helped in clarifying the paper
\end{acks}

\section*{Availability}

The code and analysis in this work is in an open source repository:
\url{https://github.com/aberke/floc-analysis}.

\bibliographystyle{ACM-Reference-Format}
\bibliography{references}

\AtEndDocument{\pagebreak
\appendix

\section{Appendix}
\label{appendix}
\setcounter{figure}{0}
\setcounter{table}{0}
\counterwithin{figure}{section}
\counterwithin{table}{section}

\subsection{comScore Data Demographics Compared to Census Estimates}

Before preprocessing we compare the demographics reported in the comScore data to census population estimates from the same time period (2017) from the U.S. Census Bureau American Community Survey (ACS) \cite{ACS:5YR} and Current Population Survey (CPS) \cite{bureauCurrentPopulationSurvey}.
We compare estimates at the household level.
These comparisons are shown in Figures~\ref{fig:comscore_demographics_state_pop} and \ref{fig:comscore_demographics_income_race}.  
When comparing the distribution of the comScore population by U.S. state there is a Pearson correlation of  0.988 (p=0.000). When comparing racial background and household income groups there is a Pearson correlation of  0.979 (p = 0.021) and 0.971 (p=0.029), respectively.
Note that even though the data are highly correlated with U.S. demographics with respect to racial background and household income, 
we create data panels that are even more representative of the U.S. population for use in our t-closeness analyses that evaluate the relationships between demographics and cohort groupings. The data panels are created using  stratified random sampling (without replacement), where strata are defined by the joint distribution of the racial background and household income demographic groups.

\begin{figure}
    \centering
    \begin{subfigure}{0.4\textwidth}
        \includegraphics[width=\textwidth]{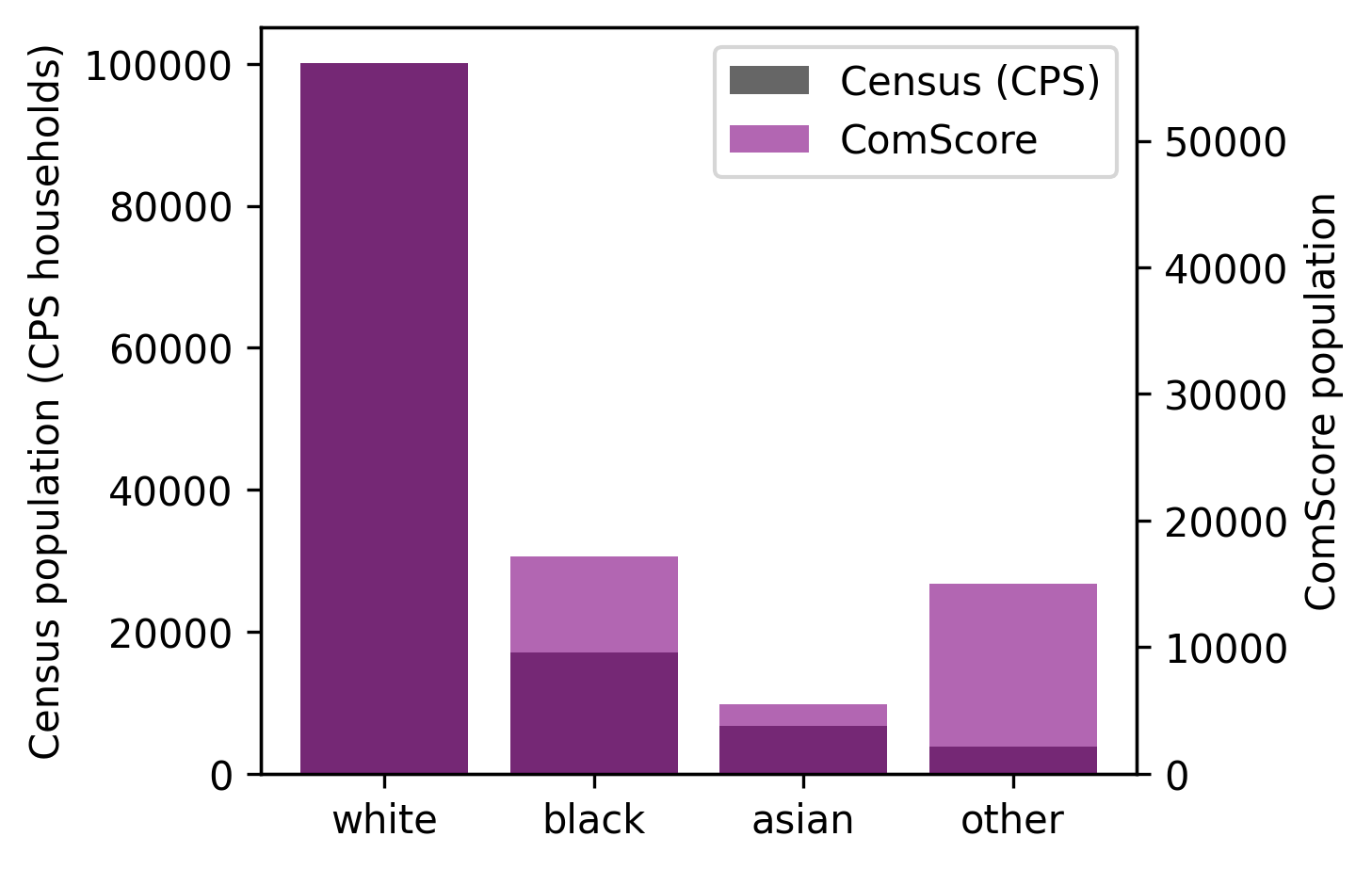}
    \end{subfigure}
    \begin{subfigure}[b]{0.4\textwidth}
        \includegraphics[width=\textwidth]{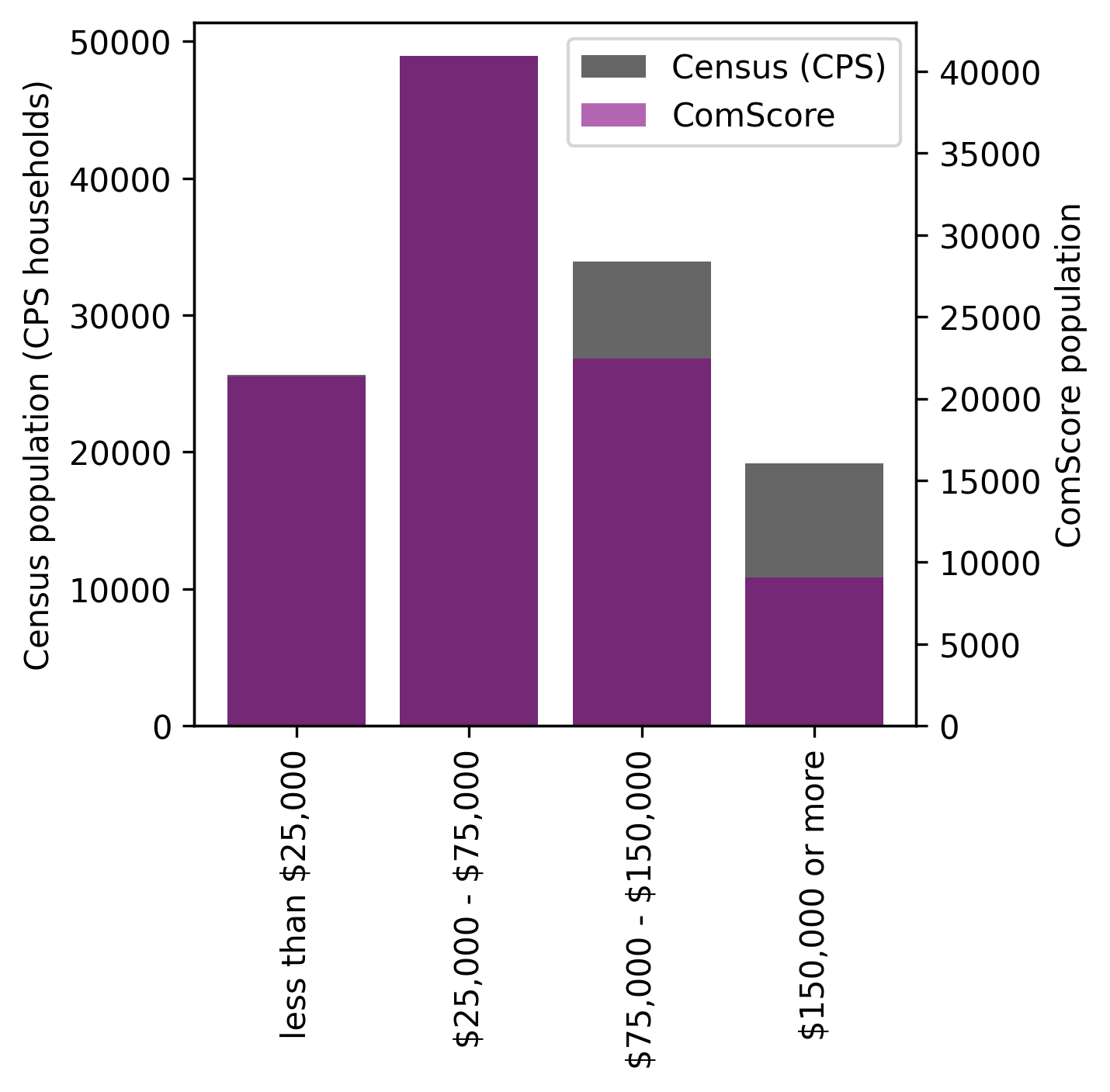}
    \end{subfigure}
    \caption{Household income and racial background demographics for the comScore versus census data from the same period (CPS 2017 \cite{bureauCurrentPopulationSurvey}).}
    \label{fig:comscore_demographics_income_race}
\end{figure}

\begin{figure*}
    \centering
    \includegraphics[width=0.9\textwidth]{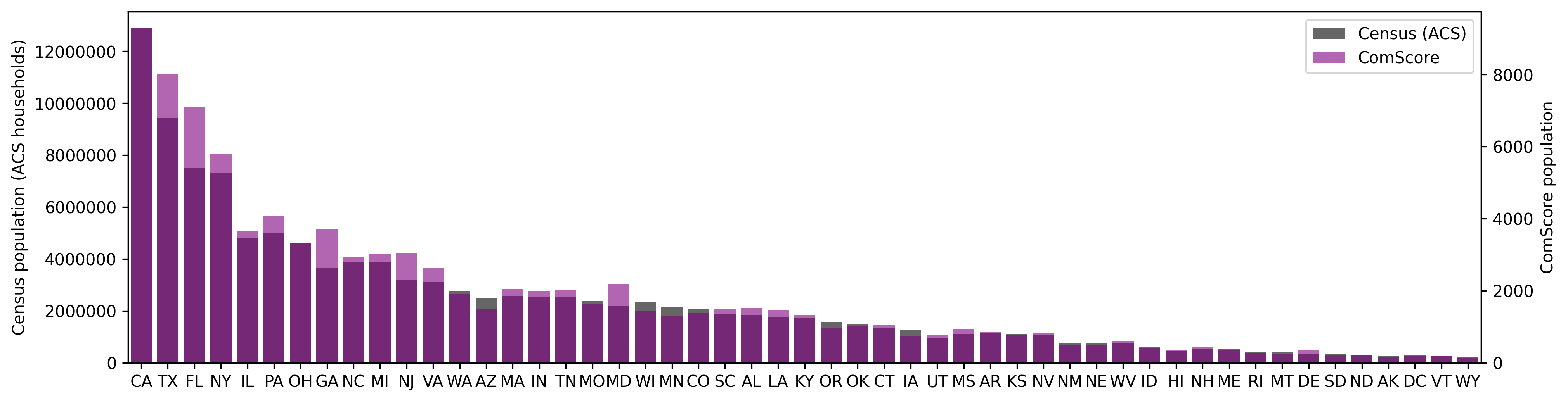}
    \caption{State population for the comScore versus census data from the same period (CPS 2017 \cite{bureauCurrentPopulationSurvey}).}
    \label{fig:comscore_demographics_state_pop}
\end{figure*}

\subsection{Distribution Of The Unique Set Of Domains Per Machine, Per Week}
\label{appendix:dist of domains per week}

Figure \ref{fig:dist_domains_per_week} shows a box plot of unique domains per machine for each week in our dataset (shown up to 50). Overall the median number of domains per week is less than the 7-domain cutoff, shown by a dashed horizontal line

\begin{figure*}
\begin{center}
\includegraphics[width=0.9\textwidth]{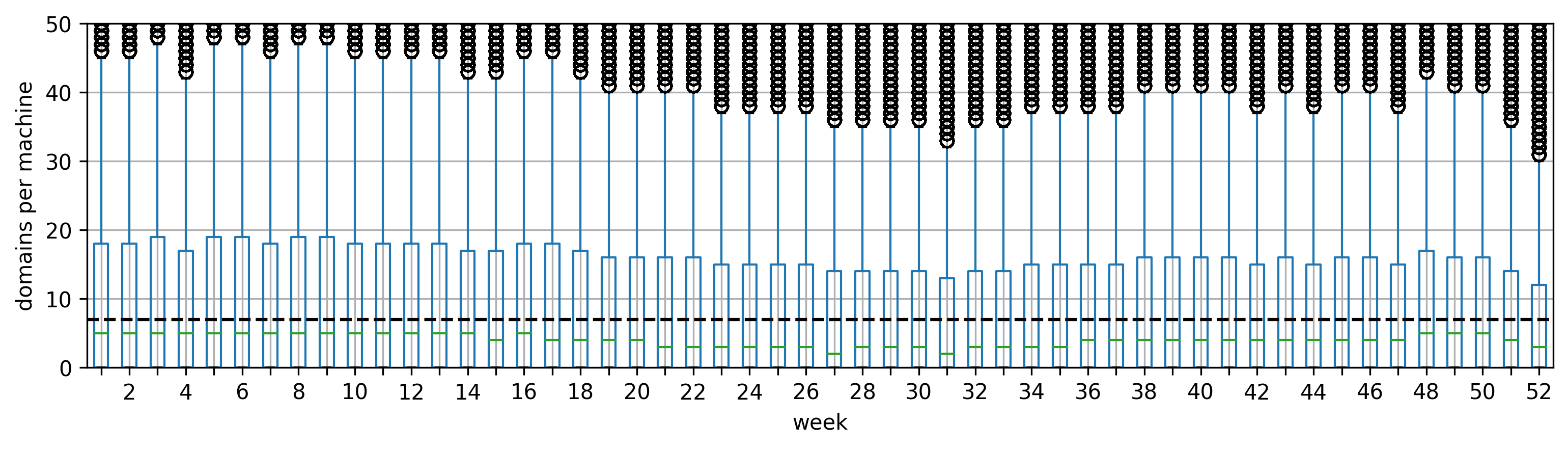}
\end{center}
\caption{
\label{fig:dist_domains_per_week}
Box plot of unique domains per machine for each week in our dataset (shown up to 50). Overall the median number of domains per week is less than the 7-domain cutoff, shown by a dashed horizontal line. 
}
\end{figure*}

\subsection{Domain Visit Frequencies By Race And Income}
\label{appendix:domains frequencies}

In order to satisfy the comScore data terms of use, we avoid showing web traffic data connected to named domains.

Table \ref{table:top100_domains_frequencies_race_income} shows the relative visit frequencies for the top 100 domains, for the aggregated overall population and by subpopulations partitioned by racial background, and then by household income group. These data represent those used in the Chi-Squared tests described in Section~\ref{section:browsing behavior differences} that are used to test browsing differences between demographic groups. Domains are ordered by their frequency in the overall population, in descending order. Domain names are not shown.
Table~\ref{table:top100_domains_names} shows the names of the top D=100 domains in alphabetical order.

\onecolumn
\begin{multicols}{2}
\end{multicols} 
\begin{scriptsize}
\begin{longtable}[c]{c|cc|cccc|cccc}
\caption{Relative visit frequencies for the top 100 domains, for the overall population and by subpopulations partitioned by demographic group. Values for the randomized comparison group used in the Chi-squared tests are also shown. Domains are ordered by their frequency in the overall population, in descending order.\label{table:top100_domains_frequencies_race_income}}\\
\hline
domain & overall & \makecell{random\\control} &  white &  black &  asian & other & \makecell{less than\\ \$25,000} & \makecell{\$25,000 -\\ \$75,000} & \makecell{\$75,000 -\\ \$150,000} & \makecell{\$150,000\\or more} \\
\hline
1   &   2.32\% &          2.31\% &  2.20\% &  2.44\% &  2.52\% &  2.75\% &             2.36\% &              2.20\% &               2.42\% &             2.54\% \\
2   &   1.54\% &          1.54\% &  1.49\% &  1.76\% &  1.45\% &  1.63\% &             1.64\% &              1.52\% &               1.55\% &             1.43\% \\
3   &   1.54\% &          1.52\% &  1.50\% &  1.87\% &  1.07\% &  1.65\% &             1.60\% &              1.53\% &               1.55\% &             1.41\% \\
4   &   1.51\% &          1.50\% &  1.49\% &  1.72\% &  1.13\% &  1.60\% &             1.49\% &              1.50\% &               1.56\% &             1.42\% \\
5   &   1.38\% &          1.37\% &  1.40\% &  1.35\% &  1.23\% &  1.42\% &             1.53\% &              1.37\% &               1.36\% &             1.19\% \\
6   &   1.26\% &          1.25\% &  1.09\% &  1.48\% &  1.62\% &  1.65\% &             1.45\% &              1.18\% &               1.18\% &             1.35\% \\
7   &   1.11\% &          1.10\% &  1.05\% &  1.31\% &  0.95\% &  1.30\% &             1.17\% &              1.09\% &               1.10\% &             1.11\% \\
8   &   0.85\% &          0.84\% &  0.89\% &  0.67\% &  0.90\% &  0.82\% &             0.74\% &              0.81\% &               0.98\% &             0.96\% \\
9   &   0.37\% &          0.37\% &  0.40\% &  0.27\% &  0.34\% &  0.37\% &             0.35\% &              0.37\% &               0.41\% &             0.34\% \\
10  &   0.34\% &          0.33\% &  0.32\% &  0.38\% &  0.30\% &  0.38\% &             0.35\% &              0.33\% &               0.34\% &             0.33\% \\
11  &   0.31\% &          0.31\% &  0.31\% &  0.33\% &  0.27\% &  0.33\% &             0.33\% &              0.31\% &               0.30\% &             0.30\% \\
12  &   0.30\% &          0.30\% &  0.34\% &  0.24\% &  0.18\% &  0.26\% &             0.23\% &              0.30\% &               0.36\% &             0.33\% \\
13  &   0.29\% &          0.29\% &  0.29\% &  0.22\% &  0.45\% &  0.30\% &             0.27\% &              0.27\% &               0.33\% &             0.39\% \\
14  &   0.28\% &          0.28\% &  0.27\% &  0.30\% &  0.26\% &  0.33\% &             0.33\% &              0.26\% &               0.27\% &             0.29\% \\
15  &   0.26\% &          0.26\% &  0.29\% &  0.19\% &  0.19\% &  0.27\% &             0.28\% &              0.27\% &               0.27\% &             0.19\% \\
16  &   0.25\% &          0.25\% &  0.27\% &  0.23\% &  0.20\% &  0.23\% &             0.27\% &              0.26\% &               0.26\% &             0.18\% \\
17  &   0.25\% &          0.25\% &  0.27\% &  0.22\% &  0.21\% &  0.24\% &             0.22\% &              0.25\% &               0.29\% &             0.26\% \\
18  &   0.25\% &          0.25\% &  0.26\% &  0.22\% &  0.22\% &  0.24\% &             0.21\% &              0.24\% &               0.29\% &             0.24\% \\
19  &   0.24\% &          0.24\% &  0.26\% &  0.18\% &  0.20\% &  0.24\% &             0.22\% &              0.24\% &               0.27\% &             0.24\% \\
20  &   0.22\% &          0.22\% &  0.20\% &  0.31\% &  0.21\% &  0.25\% &             0.26\% &              0.22\% &               0.21\% &             0.19\% \\
21  &   0.21\% &          0.21\% &  0.15\% &  0.37\% &  0.28\% &  0.34\% &             0.27\% &              0.19\% &               0.18\% &             0.28\% \\
22  &   0.20\% &          0.20\% &  0.20\% &  0.23\% &  0.21\% &  0.20\% &             0.19\% &              0.19\% &               0.22\% &             0.23\% \\
23  &   0.20\% &          0.20\% &  0.19\% &  0.19\% &  0.24\% &  0.21\% &             0.18\% &              0.18\% &               0.21\% &             0.27\% \\
24  &   0.20\% &          0.20\% &  0.19\% &  0.24\% &  0.21\% &  0.22\% &             0.23\% &              0.21\% &               0.19\% &             0.17\% \\
25  &   0.18\% &          0.18\% &  0.19\% &  0.13\% &  0.24\% &  0.17\% &             0.13\% &              0.17\% &               0.23\% &             0.21\% \\
26  &   0.17\% &          0.17\% &  0.16\% &  0.10\% &  0.21\% &  0.25\% &             0.13\% &              0.14\% &               0.16\% &             0.38\% \\
27  &   0.17\% &          0.17\% &  0.16\% &  0.21\% &  0.11\% &  0.17\% &             0.20\% &              0.17\% &               0.15\% &             0.12\% \\
28  &   0.16\% &          0.16\% &  0.16\% &  0.13\% &  0.17\% &  0.17\% &             0.13\% &              0.15\% &               0.18\% &             0.15\% \\
29  &   0.15\% &          0.15\% &  0.13\% &  0.25\% &  0.13\% &  0.15\% &             0.18\% &              0.15\% &               0.14\% &             0.11\% \\
30  &   0.15\% &          0.16\% &  0.14\% &  0.12\% &  0.28\% &  0.20\% &             0.15\% &              0.13\% &               0.15\% &             0.26\% \\
31  &   0.15\% &          0.15\% &  0.15\% &  0.17\% &  0.16\% &  0.17\% &             0.16\% &              0.15\% &               0.16\% &             0.16\% \\
32  &   0.15\% &          0.15\% &  0.15\% &  0.12\% &  0.25\% &  0.15\% &             0.11\% &              0.14\% &               0.19\% &             0.20\% \\
33  &   0.14\% &          0.14\% &  0.15\% &  0.14\% &  0.08\% &  0.12\% &             0.11\% &              0.14\% &               0.16\% &             0.15\% \\
34  &   0.14\% &          0.14\% &  0.13\% &  0.16\% &  0.15\% &  0.14\% &             0.13\% &              0.13\% &               0.16\% &             0.17\% \\
35  &   0.14\% &          0.13\% &  0.12\% &  0.16\% &  0.16\% &  0.15\% &             0.14\% &              0.13\% &               0.14\% &             0.15\% \\
36  &   0.14\% &          0.14\% &  0.14\% &  0.12\% &  0.13\% &  0.15\% &             0.14\% &              0.14\% &               0.13\% &             0.13\% \\
37  &   0.14\% &          0.14\% &  0.14\% &  0.12\% &  0.12\% &  0.14\% &             0.14\% &              0.14\% &               0.14\% &             0.14\% \\
38  &   0.13\% &          0.13\% &  0.14\% &  0.12\% &  0.10\% &  0.12\% &             0.13\% &              0.14\% &               0.13\% &             0.10\% \\
39  &   0.13\% &          0.13\% &  0.13\% &  0.14\% &  0.12\% &  0.12\% &             0.14\% &              0.12\% &               0.13\% &             0.13\% \\
40  &   0.13\% &          0.13\% &  0.13\% &  0.07\% &  0.17\% &  0.20\% &             0.09\% &              0.11\% &               0.13\% &             0.32\% \\
41  &   0.13\% &          0.13\% &  0.14\% &  0.13\% &  0.08\% &  0.12\% &             0.11\% &              0.13\% &               0.15\% &             0.15\% \\
42  &   0.12\% &          0.12\% &  0.13\% &  0.11\% &  0.12\% &  0.12\% &             0.10\% &              0.12\% &               0.14\% &             0.13\% \\
43  &   0.12\% &          0.12\% &  0.13\% &  0.11\% &  0.08\% &  0.11\% &             0.10\% &              0.12\% &               0.14\% &             0.10\% \\
44  &   0.12\% &          0.12\% &  0.12\% &  0.08\% &  0.18\% &  0.13\% &             0.10\% &              0.10\% &               0.13\% &             0.20\% \\
45  &   0.12\% &          0.12\% &  0.13\% &  0.09\% &  0.12\% &  0.11\% &             0.10\% &              0.12\% &               0.15\% &             0.15\% \\
46  &   0.11\% &          0.11\% &  0.10\% &  0.11\% &  0.13\% &  0.12\% &             0.11\% &              0.10\% &               0.11\% &             0.14\% \\
47  &   0.11\% &          0.10\% &  0.11\% &  0.07\% &  0.13\% &  0.10\% &             0.07\% &              0.09\% &               0.14\% &             0.16\% \\
48  &   0.11\% &          0.11\% &  0.09\% &  0.16\% &  0.09\% &  0.16\% &             0.14\% &              0.11\% &               0.10\% &             0.11\% \\
49  &   0.11\% &          0.11\% &  0.11\% &  0.13\% &  0.14\% &  0.13\% &             0.12\% &              0.11\% &               0.11\% &             0.14\% \\
50  &   0.11\% &          0.11\% &  0.11\% &  0.08\% &  0.16\% &  0.10\% &             0.08\% &              0.09\% &               0.14\% &             0.17\% \\
51  &   0.11\% &          0.10\% &  0.10\% &  0.12\% &  0.10\% &  0.11\% &             0.13\% &              0.11\% &               0.09\% &             0.08\% \\
52  &   0.11\% &          0.11\% &  0.12\% &  0.09\% &  0.09\% &  0.09\% &             0.08\% &              0.10\% &               0.13\% &             0.16\% \\
53  &   0.11\% &          0.11\% &  0.11\% &  0.12\% &  0.09\% &  0.12\% &             0.10\% &              0.11\% &               0.12\% &             0.11\% \\
54  &   0.11\% &          0.11\% &  0.13\% &  0.07\% &  0.09\% &  0.10\% &             0.08\% &              0.11\% &               0.14\% &             0.11\% \\
55  &   0.10\% &          0.10\% &  0.11\% &  0.08\% &  0.06\% &  0.08\% &             0.08\% &              0.10\% &               0.12\% &             0.11\% \\
56  &   0.10\% &          0.10\% &  0.10\% &  0.06\% &  0.14\% &  0.10\% &             0.08\% &              0.09\% &               0.12\% &             0.13\% \\
57  &   0.10\% &          0.10\% &  0.10\% &  0.10\% &  0.09\% &  0.10\% &             0.10\% &              0.09\% &               0.10\% &             0.11\% \\
58  &   0.10\% &          0.09\% &  0.08\% &  0.13\% &  0.15\% &  0.11\% &             0.11\% &              0.08\% &               0.10\% &             0.12\% \\
59  &   0.10\% &          0.10\% &  0.10\% &  0.10\% &  0.09\% &  0.10\% &             0.10\% &              0.10\% &               0.09\% &             0.09\% \\
60  &   0.09\% &          0.09\% &  0.09\% &  0.06\% &  0.09\% &  0.08\% &             0.06\% &              0.08\% &               0.11\% &             0.12\% \\
61  &   0.09\% &          0.09\% &  0.09\% &  0.06\% &  0.07\% &  0.08\% &             0.07\% &              0.08\% &               0.10\% &             0.11\% \\
62  &   0.09\% &          0.09\% &  0.08\% &  0.07\% &  0.15\% &  0.10\% &             0.08\% &              0.08\% &               0.09\% &             0.15\% \\
63  &   0.09\% &          0.09\% &  0.09\% &  0.09\% &  0.09\% &  0.09\% &             0.09\% &              0.09\% &               0.09\% &             0.09\% \\
64  &   0.09\% &          0.09\% &  0.09\% &  0.10\% &  0.09\% &  0.10\% &             0.10\% &              0.09\% &               0.09\% &             0.10\% \\
65  &   0.09\% &          0.09\% &  0.09\% &  0.10\% &  0.09\% &  0.10\% &             0.09\% &              0.09\% &               0.09\% &             0.09\% \\
66  &   0.09\% &          0.09\% &  0.09\% &  0.12\% &  0.06\% &  0.09\% &             0.09\% &              0.10\% &               0.10\% &             0.08\% \\
67  &   0.09\% &          0.09\% &  0.10\% &  0.04\% &  0.07\% &  0.06\% &             0.06\% &              0.08\% &               0.11\% &             0.10\% \\
68  &   0.09\% &          0.09\% &  0.08\% &  0.11\% &  0.13\% &  0.10\% &             0.10\% &              0.07\% &               0.09\% &             0.11\% \\
69  &   0.08\% &          0.08\% &  0.08\% &  0.07\% &  0.11\% &  0.08\% &             0.07\% &              0.08\% &               0.09\% &             0.09\% \\
70  &   0.08\% &          0.08\% &  0.07\% &  0.09\% &  0.08\% &  0.08\% &             0.08\% &              0.08\% &               0.08\% &             0.08\% \\
71  &   0.08\% &          0.08\% &  0.08\% &  0.08\% &  0.07\% &  0.08\% &             0.09\% &              0.09\% &               0.08\% &             0.07\% \\
72  &   0.08\% &          0.08\% &  0.08\% &  0.07\% &  0.07\% &  0.07\% &             0.09\% &              0.08\% &               0.07\% &             0.06\% \\
73  &   0.08\% &          0.08\% &  0.08\% &  0.07\% &  0.11\% &  0.09\% &             0.08\% &              0.08\% &               0.09\% &             0.09\% \\
74  &   0.08\% &          0.08\% &  0.09\% &  0.08\% &  0.07\% &  0.07\% &             0.07\% &              0.09\% &               0.09\% &             0.07\% \\
75  &   0.08\% &          0.08\% &  0.08\% &  0.05\% &  0.08\% &  0.07\% &             0.05\% &              0.06\% &               0.09\% &             0.12\% \\
76  &   0.08\% &          0.08\% &  0.09\% &  0.03\% &  0.04\% &  0.06\% &             0.05\% &              0.07\% &               0.10\% &             0.10\% \\
77  &   0.07\% &          0.07\% &  0.06\% &  0.07\% &  0.09\% &  0.08\% &             0.07\% &              0.06\% &               0.07\% &             0.08\% \\
78  &   0.07\% &          0.07\% &  0.06\% &  0.10\% &  0.06\% &  0.06\% &             0.08\% &              0.07\% &               0.07\% &             0.06\% \\
79  &   0.07\% &          0.07\% &  0.08\% &  0.05\% &  0.06\% &  0.06\% &             0.06\% &              0.07\% &               0.08\% &             0.07\% \\
80  &   0.07\% &          0.07\% &  0.07\% &  0.04\% &  0.06\% &  0.06\% &             0.05\% &              0.06\% &               0.08\% &             0.10\% \\
81  &   0.07\% &          0.07\% &  0.08\% &  0.07\% &  0.05\% &  0.06\% &             0.07\% &              0.08\% &               0.07\% &             0.05\% \\
82  &   0.07\% &          0.07\% &  0.08\% &  0.05\% &  0.05\% &  0.06\% &             0.06\% &              0.07\% &               0.07\% &             0.06\% \\
83  &   0.07\% &          0.07\% &  0.07\% &  0.10\% &  0.06\% &  0.08\% &             0.10\% &              0.07\% &               0.06\% &             0.06\% \\
84  &   0.07\% &          0.07\% &  0.07\% &  0.06\% &  0.04\% &  0.08\% &             0.07\% &              0.07\% &               0.07\% &             0.07\% \\
85  &   0.07\% &          0.07\% &  0.05\% &  0.12\% &  0.06\% &  0.07\% &             0.08\% &              0.06\% &               0.06\% &             0.05\% \\
86  &   0.07\% &          0.07\% &  0.08\% &  0.04\% &  0.06\% &  0.06\% &             0.05\% &              0.07\% &               0.09\% &             0.07\% \\
87  &   0.07\% &          0.07\% &  0.09\% &  0.04\% &  0.04\% &  0.06\% &             0.05\% &              0.07\% &               0.09\% &             0.07\% \\
88  &   0.07\% &          0.07\% &  0.06\% &  0.07\% &  0.11\% &  0.07\% &             0.05\% &              0.06\% &               0.09\% &             0.09\% \\
89  &   0.07\% &          0.07\% &  0.08\% &  0.04\% &  0.04\% &  0.05\% &             0.05\% &              0.07\% &               0.08\% &             0.07\% \\
90  &   0.07\% &          0.07\% &  0.07\% &  0.06\% &  0.09\% &  0.07\% &             0.05\% &              0.06\% &               0.09\% &             0.08\% \\
91  &   0.07\% &          0.07\% &  0.09\% &  0.04\% &  0.05\% &  0.06\% &             0.06\% &              0.07\% &               0.08\% &             0.08\% \\
92  &   0.07\% &          0.07\% &  0.07\% &  0.08\% &  0.05\% &  0.07\% &             0.07\% &              0.07\% &               0.07\% &             0.06\% \\
93  &   0.07\% &          0.07\% &  0.07\% &  0.04\% &  0.13\% &  0.06\% &             0.04\% &              0.06\% &               0.09\% &             0.09\% \\
94  &   0.07\% &          0.07\% &  0.07\% &  0.06\% &  0.05\% &  0.06\% &             0.06\% &              0.07\% &               0.08\% &             0.08\% \\
95  &   0.07\% &          0.07\% &  0.05\% &  0.11\% &  0.12\% &  0.10\% &             0.10\% &              0.06\% &               0.06\% &             0.06\% \\
96  &   0.07\% &          0.07\% &  0.07\% &  0.08\% &  0.07\% &  0.07\% &             0.08\% &              0.07\% &               0.08\% &             0.07\% \\
97  &   0.07\% &          0.07\% &  0.07\% &  0.07\% &  0.06\% &  0.07\% &             0.07\% &              0.07\% &               0.06\% &             0.05\% \\
98  &   0.06\% &          0.06\% &  0.06\% &  0.06\% &  0.07\% &  0.07\% &             0.05\% &              0.06\% &               0.07\% &             0.09\% \\
99  &   0.06\% &          0.06\% &  0.05\% &  0.11\% &  0.05\% &  0.08\% &             0.08\% &              0.06\% &               0.06\% &             0.05\% \\
100 &   0.06\% &          0.06\% &  0.07\% &  0.04\% &  0.04\% &  0.05\% &             0.05\% &              0.06\% &               0.08\% &             0.07\% \\
\hline
\end{longtable}
\end{scriptsize}

\begin{table*}[h]
\caption{Top 100 domain names in alphabetical order.}
\scriptsize
\begin{minipage}[t]{0.24\textwidth}
\begin{tabular}{ll}
\hline
& domain \\
\hline
1   &            247-inc.net \\
2   &              adobe.com \\
3   &                adp.com \\
4   &             amazon.com \\
5   &                aol.com \\
6   &              apple.com \\
7   &                ask.com \\
8   &                att.com \\
9   &      bangcreatives.com \\
10  &      bankofamerica.com \\
11  &            bestbuy.com \\
12  &               bing.com \\
13  &          bongacams.com \\
14  &              btrll.com \\
15  &           camdolls.com \\
16  &         capitalone.com \\
17  &              chase.com \\
18  &         chaturbate.com \\
19  &               citi.com \\
20  &                cnn.com \\
21  &            comcast.net \\
22  &         craigslist.org \\
23  &            dropbox.com \\
24  &               ebay.com \\
25  &     elbowviewpoint.com \\
\vdots & \vdots \\ 
\vdots & \vdots \\ 
\end{tabular}
\end{minipage} \hfill
\begin{minipage}[t]{0.25\textwidth}
\begin{tabular}{ll}
\hline
& domain \\
\hline
\vdots & \vdots \\

26  &               espn.com \\
27  &               etsy.com \\
28  &            expedia.com \\
29  &           facebook.com \\
30  &            foxnews.com \\
31  &                 go.com \\
32  &             google.com \\
33  &  googlesyndication.com \\
34  &          homedepot.com \\
35  &               hulu.com \\
36  &            ibtimes.com \\
37  &               imdb.com \\
38  &             indeed.com \\
39  &          instagram.com \\
40  &        instructure.com \\
41  &             intuit.com \\
42  &              kohls.com \\
43  &           linkedin.com \\
44  &               live.com \\
45  &         livejasmin.com \\
46  &              lowes.com \\
47  &              macys.com \\
48  &           mapquest.com \\
49  &             mcafee.com \\
50  &          microsoft.com \\
\vdots & \vdots \\ 
\end{tabular}
\end{minipage} \hfill
\begin{minipage}[t]{0.24\textwidth}
\begin{tabular}{ll}
\hline
& domain \\
\hline
\vdots & \vdots \\
51  &    microsoftonline.com \\
52  &            mozilla.org \\
53  &                msn.com \\
54  &              myway.com \\
55  &            netflix.com \\
56  &             norton.com \\
57  &            nytimes.com \\
58  &             office.com \\
59  &            pandora.com \\
60  &             paypal.com \\
61  &          pinterest.com \\
62  &            pornhub.com \\
63  &        pornhublive.com \\
64  &            quizlet.com \\
65  &            realtor.com \\
66  &             reddit.com \\
67  &            redtube.com \\
68  &          reference.com \\
69  &             roblox.com \\
70  &           s3xified.com \\
71  &            safesear.ch \\
72  &              skype.com \\
73  &      smartadserver.com \\
74  &     steamcommunity.com \\
75  &       steampowered.com \\
\vdots & \vdots \\ 
\end{tabular}
\end{minipage} \hfill
\begin{minipage}[t]{0.24\textwidth}
\begin{tabular}{ll}
\hline
& domain \\
\hline
\vdots & \vdots \\ 
\vdots & \vdots \\
76  &              taleo.net \\
77  &             target.com \\
78  &        tripadvisor.com \\
79  &             tumblr.com \\
80  &            twitter.com \\
81  &    verizonwireless.com \\
82  &            walmart.com \\
83  &     washingtonpost.com \\
84  &            weather.com \\
85  &              webmd.com \\
86  &         wellsfargo.com \\
87  &              wikia.com \\
88  &            wikihow.com \\
89  &          wikipedia.org \\
90  &          wordpress.com \\
91  &            xfinity.com \\
92  &           xhamster.com \\
93  &       xhamsterlive.com \\
94  &               xnxx.com \\
95  &            xvideos.com \\
96  &              yahoo.com \\
97  &               yelp.com \\
98  &            youporn.com \\
99  &            youtube.com \\
100 &             zillow.com \\
\hline
\end{tabular}
\end{minipage}

\label{table:top100_domains_names}
\end{table*}

\begin{multicols}{2}

\subsection{Baseline t-closeness Analysis With U.S. Population}

\begin{figure*}
    \centering
    \begin{subfigure}{\textwidth}
        \includegraphics[width=\textwidth]{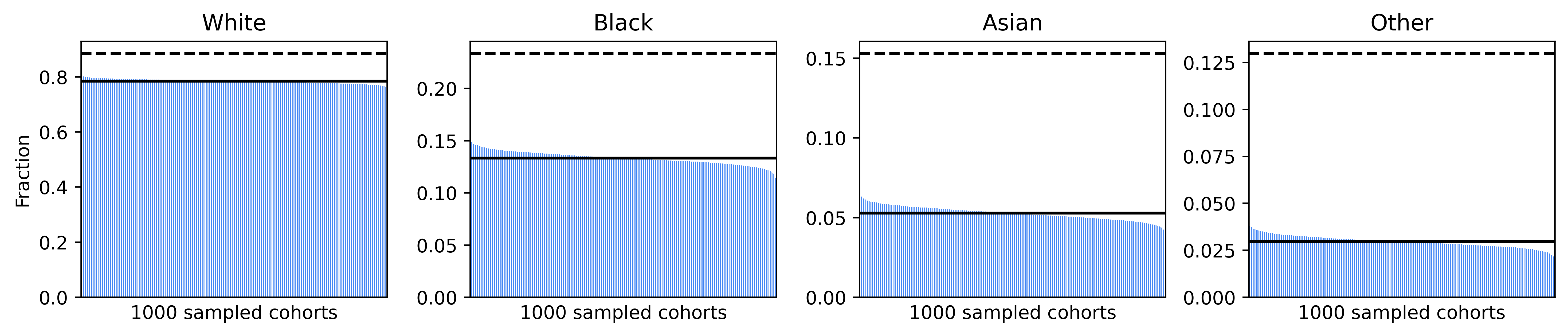}
    \end{subfigure}
    \begin{subfigure}{\textwidth}
        \includegraphics[width=\textwidth]{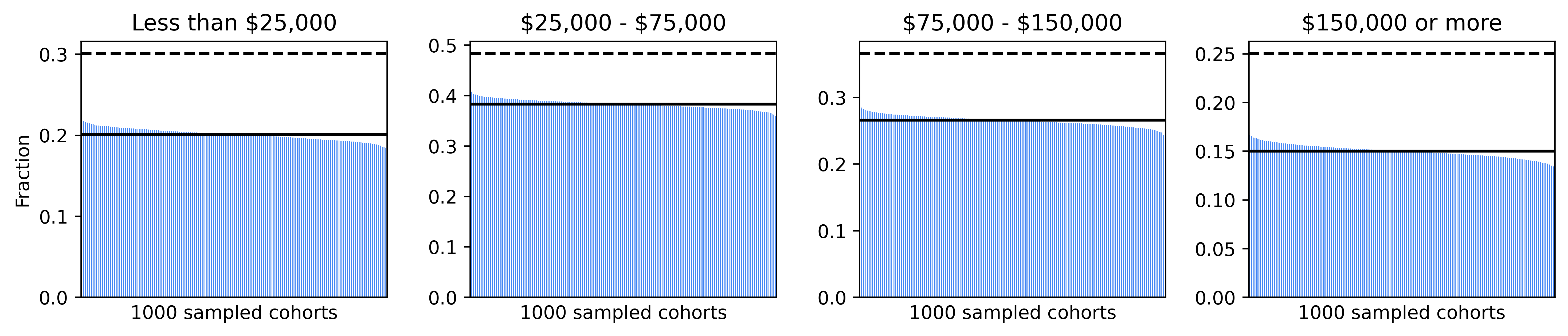}
    \end{subfigure}
\caption{
\label{fig:us_pop_t_closeness}
t-closeness analysis for OT sized panel and U.S. population.
For each cohort, we check for each demographic group whether the fraction of users in that demographic group exceeds the population mean by more than $t=0.1$. For illustration, we show this analysis for 1000 randomly sampled cohorts. The solid black line indicates the fraction of each demographic group in the population. The dashed line indicates the threshold where t-closeness for $t=0.1$ would be violated.
}
\end{figure*}

For baseline t-closeness analysis we create a random panel matching the estimated size of the FLoC OT, where user demographic groups and cohort IDs are randomly assigned. We assign the household income groups and racial background groups so that their joint distribution matches the U.S. population estimates (CPS 2017 \cite{bureauCurrentPopulationSurvey}).
To estimate the size of the FLoC OT, we use our data to compute cohorts at varying levels of $k$ and find a consistent relationship where (mean cohort size) / $k$ $\sim$ 1.5. The FLoC OT had $k=2000$ and 33,872 cohorts.
From this data we estimate the FLoC OT size with mean cohort sizes of 3000 (k=2000 x 1.5) and 101,616,000 (3000 x 33,872) user devices.
We assign the 33,872 cohort IDs to the 101,616,000 users by randomly assigning the first 2000 for each cohort and then using a uniform distribution over cohort IDs for the remaining assignments.
We then use this panel to do the following t-closeness check for the household income groups and racial background groups separately.
For each cohort, we check for each demographic group whether the fraction of users in that demographic group exceeds the fraction in the general population by more than $t=0.1$. No cohorts exceed this threshold. This is illustrated in Figure~\ref{fig:us_pop_t_closeness}.

\end{multicols}
} 

\end{document}